\documentclass[nofootinbib,superscriptaddress,twocolumn]{revtex4-2}
\usepackage[a4paper,left=2.0cm,top=2.0cm,right=1.0cm,bottom=2.0cm]{geometry}
\usepackage{amssymb,amsmath,amsthm,graphicx,cancel,braket,comment,bm,caption,subcaption,float,cancel,braket,dsfont,graphicx}
\usepackage{mathrsfs,hyperref,subcaption,xcolor}
\usepackage{newtxtext,newtxmath}
\hypersetup{final,colorlinks=true,	linkcolor={blue}, citecolor={red}, 	urlcolor={violet}}

\newcommand{\at}{\dot{a}}
\newcommand{\att}{\ddot{a}}
\newcommand{\Nt}{\dot{N}}

\newcommand{\p}{\partial}
\newcommand{\la}{\lambda}

\begin{document}
\title{\textbf{Quantum geometric inflation}}
\author{Nephtalí Eliceo Martínez Pérez \thanks{mp218570084@egresados.buap.mx}}
\author{Cupatitzio Ramírez Romero \thanks{cramirez@fcfm.buap.mx}}
\affiliation{Facultad de Ciencias Físico Matemáticas, Benemérita Universidad Autónoma de Puebla\\
Avenida San Claudio y 18 Sur, Puebla, Puebla, México}
	
\begin{abstract}
	Geometric Inflation adds an infinite tower of curvature invariants to the Einstein-Hilbert action while preserving certain key features such as leading to second order modified Friedmann equations for FRW backgrounds. Sourcing them with a perfect barotropic fluid can easily produce a slow-roll inflationary regime, no inflaton field is required. In this work, we perform the Wheeler-DeWitt quantization of minisuperspace Geometric Inflation models. First, we construct an effective 1D FRW action and perform the Hamiltonian formulation using both Ostrogradsky's approach and Dirac's algorithm for constrained systems. The Lagrangian and Hamiltonian treatment of the perfect fluid is based on Schutz's formalism.  Second, we postulate the Wheeler-DeWitt equation and investigate its solutions both analytically and numerically. For radiation we are led to a system of fractional differential equations for which we use Caputo fractional derivative operator. We construct wave packets in momentum space, with the fluid's variable acting as external time parameter.
\end{abstract}
	
\maketitle
\section{Introduction}
In the last few years, a class of modified gravity models, called collectively Generalized quasi-topological gravity (GQTG), has been the subject of intense research \cite{Hennigar, Bueno2019, Bueno_2020, Bueno_2023, PhysRevD.108.044016}. They are higher-curvature generalizations of Einstein's gravity in $D$-dimensions that exhibit a number of nontrivial features. In particular, the field equations linearized around maximally symmetric backgrounds are second order and propagate only a transverse and traceless graviton. On the other hand, when evaluated on a general static spherically symmetric ansatz, the equations of motion admit non-hairy generalizations of Schwarzschild’s black hole solution characterized by a single function $g_{tt}(r)=g_{rr}^{-1}(r)$.

GQTG has also been considered in cosmology to address the late-time accelerated cosmic expansion and the presumed inflationary period of the very early universe. As first demonstrated with the cubic case \cite{ARCINIEGA2020135272}, when evaluated on an FRW ansatz, certain GQGTs yield also second order equations of motion. The new Friedmann equations are modified in a nontrivial way, allowing accelerated expansion in the presence of a barotropic perfect fluid. The cubic case was then generalized to curvature terms of fourth and fifth order in \cite{CISTERNA2020135435}. The accelerated expansion generated by models with curvature terms up to a finite order is however of power law type. For quasi exponential expansion rate, an infinite tower of curvature invariants was introduced in \cite{ARCINIEGA2020135242}, in a framework coined Geometric Inflation (GI). It was shown models not truncated arbitrarily at some power of curvature, can produce a suitable inflationary phase with graceful exit, driven by a perfect fluid such as radiation, thus avoiding scalar fields. The viability of GI as a framework for cosmological evolution was analyzed in \cite{JAIME2021100887}. The slow roll parameters are compatible with the current constraints \cite{ARCINIEGA2020135731}. It has been argued for the cubic case that a scalar field would be beneficial to avoid super-Planckian values of the initial energy density \cite{Edelstein_2020,Edelstein2021}. This setup is not restricted to inflation but can account for the subsequent history of the universe up to the current accelerated expansion \cite{universe12050129}. 

On the other hand, Quantum Cosmology (QC) constitutes an effort to shed light on fundamental questions such as the Big Bang singularity and the (classical) initial state of the universe by resorting to concepts and methods of quantum theory \cite{wiltshire2003,halliwell}. The issue of initial conditions is particularly relevant in connection with the inflationary paradigm. For instance, while single field models certainly produce a period of accelerated expansion, it still requires certain conditions  such as a sufficiently large initial field value for the universe to inflate as required to fit the current cosmological picture \cite{baumann2012tasilecturesinflation}. Therefore, it is reasonable to ask the space of initial states and its probability distribution. The output of QC, namely, the wave function of the universe can in principle be used to construct a probability distribution of initial states \cite{mijic, higherderivatives}. Now, whereas geometric inflation dispenses with an inflaton field, although it can certainly be complemented with scalar fields \cite{Edelstein2021,einsteiniancubic}, solutions depend on the initial Hubble factor or energy density of the fluid. 

To address such questions for geometric inflation, we must first work out a quantization scheme. In this work, we explore minisuperspace version of Quantum Geometrodynamics \cite{moniz,Kiefer:2004xyv}, where the Hamiltonian constraint of an FRW cosmological model is promoted to a quantum constraint $\mathcal{H}_0 \Psi=0$, the Wheeler-DeWitt (WDW) equation,  obeyed by the wave function defined in a space of finite dimension, the minisuperpace, whose points are labeled by the values of scale factor together with some homogeneous matter fields. These toy models are useful to explore technical and conceptual aspects of the different quantization approaches while offering insights into the quantum nature of the universe \cite{Kiefer:2004xyv}.

In this work, we study the Wheeler-DeWitt (WDW) quantization of spatially flat minisuperspace models of geometric inflation. The paper is organized as follows. In Section \ref{sec2}, we present the Friedman equations of geometric inflation and briefly describe the solutions for dust and radiation. In Section \ref{sec3}, we construct an effective 1D FRW action for geometric inflation guided by two requirements: (a) the equation of motion of the scale factor, in the proper time gauge, is second order, and (b) the contribution to the Friedmann constraint, from the $n$-th curvature term, is proportional to the Hubble factor to the power $2n$. Next, we perform the canonical formulation using Ostrogradsky's construction for higher-derivatives and Dirac's algorithm for constrained systems. In Section \ref{sec4}, we write the full Hamiltonian of geometric inflation plus a linear barotropic perfect fluid, using Schutz formalism for the latter. We consider the WDW equation in the position representation, which amounts to an infinite-order differential equation, and describe a WKB approximation. In Section \ref{sec5}, we write the WDW equation directly in the momentum representation and solve it for pressure free matter or dust. The wave functions are labeled by a continuous parameter, thus we construct a wave packet with which we compute the mean value of the Hubble factor, with the fluid's variable playing the role of time. In \ref{sec5.2}, we consider radiation, in which case we are led to a fractional differential WDW equation. We only use numerical solutions using solvers available in the literature. Section \ref{sec6} is dedicated to conclusions. In Appendix \ref{app1}, we briefly describe Schutz's  velocity-potential formulation of perfect fluids and its FRW reduction. Appendix \ref{app2} we detailed the integration of a special function. In Appendix \ref{app3}, we collect the basic ideas leading to formulation of fractional differential equations, following \cite{diethelm}, and the algorithms to solve them.

\section{Geometric inflation}\label{sec2}
As explained in reference \cite{ARCINIEGA2020135242}, at every order $n\in \mathbb{N}$ there are invariants $\mathcal{R}_{(n)}$, obtained by contracting the Riemann and metric tensors, satisfying the aforementioned features of GQTG and also yielding a well posed cosmological problem with second order Friedmann equations. In the supplementary material of \cite{ARCINIEGA2020135242}, authors list the explicit $\mathcal{R}_{(n)}$ up to $n=8$ and give the recipe to construct all higher orders terms, $\mathcal{R}_{(2)}$ is the Gauss-Bonnet density, thus the generic action of Geometric Inflation takes the form
\begin{align}\label{generalized}
	S=\frac{1}{2\kappa^2} \int d^4x\,  \sqrt{-g}\, \Big[R-2 \Lambda+\sum_{n=3}^\infty \la_n L^{2n-2} \mathcal{R}_{(n)}\Big],
\end{align}
where $\kappa^2=8\pi G=M_P^{-2}$, $\la_n$ are dimensionless constants and $L^{-1}$ is an energy scale large enough to make the influence of $\mathcal{R}_{(n)}$ negligible for small curvature. 

For a homogeneous and isotropic, or FRW universe, whose metric, in comoving coordinates $(t,r,\theta,\phi)$, is given by     
\begin{align}\label{frwmetric}
	ds^2=-N^2(t) dt^2+a^2(t) \left[ \frac{dr^2}{1-k r^2}+r^2 (d\theta^2+\sin^2 \theta d\phi^2)\right], 
\end{align}
the modified Friedmann and the conservation equations, in the proper time gauge $N=1$, read
\begin{subequations}\label{friedmann}
\begin{align}
3 F(h)&=\kappa^2 \rho+\Lambda, \label{frid} \\
\dot{H}&=\frac{-\kappa^2 (p+\rho)}{F'(h)/h}+\frac{k}{a^2}, \label{accele} \\
\dot{\rho}&=-3 H (p+\rho), \label{continuity}
\end{align}
\end{subequations}
where the function $h$ of two arguments is defined as follows, 
\begin{align}\label{hchikis}
	h(a,\at)\equiv \frac{\sqrt{\at^2+k}}{a}.
\end{align}
$p, \rho$ are, respectively, the pressure and energy density of the perfect fluid. 

For general relativity (GR), $F(h)=h^2$, whereas for geometric inflation,  
\begin{align}\label{Fseries}
	F(h)=h^2+L^{-2} \sum_{n=3}^\infty (-)^n \la_n (L h)^{2n}. 
\end{align} 

For a concrete model, the parameters $\la_n$ must be specified in some way. For example, choosing $\la_{2n+1}=0$ and $\la_{4+2n}=\frac{\la}{n!}$ gives rise to Model 1 of \cite{ARCINIEGA2020135242}. Setting $k=0$, $h^2=H^2$, 
\begin{align}\label{model1}
\text{Model 1:} \quad 	F(H)=H^2+\la L^6 H^8 e^{L^4 H^4},
\end{align}
On the other hand, setting $\la_3\equiv \la>0$ and $\la_{n\ge 4}=(-)^n \frac{\la}{(n-4)!}$, one obtains 
\begin{align}\label{model2}
\text{Model 2:} \quad	F(H)=H^2-\lambda [L^4 H^6-L^6 H^8 e^{L^2 H^2}].
\end{align}
We shall use these models to explore the minisuperspace quantization of Geometric Inflation.

\begin{figure*}[ht]
	\centering
	\includegraphics[width=.38\textwidth]{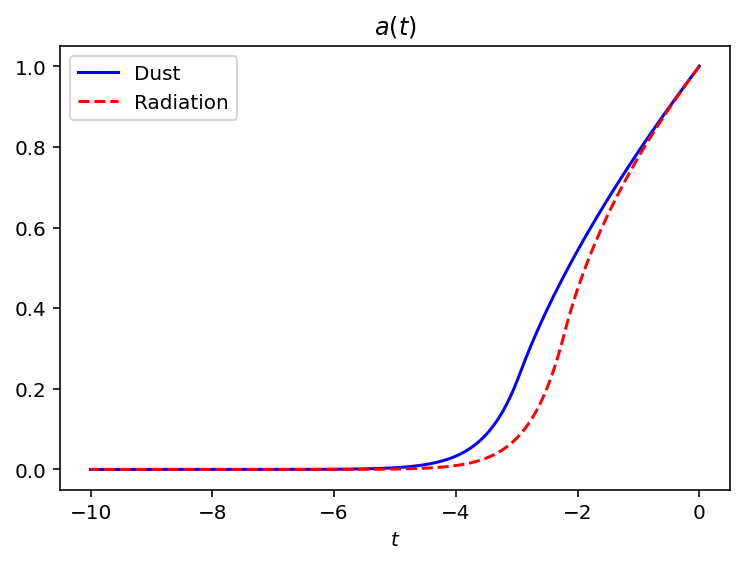}
	\qquad
	\includegraphics[width=.38\textwidth]{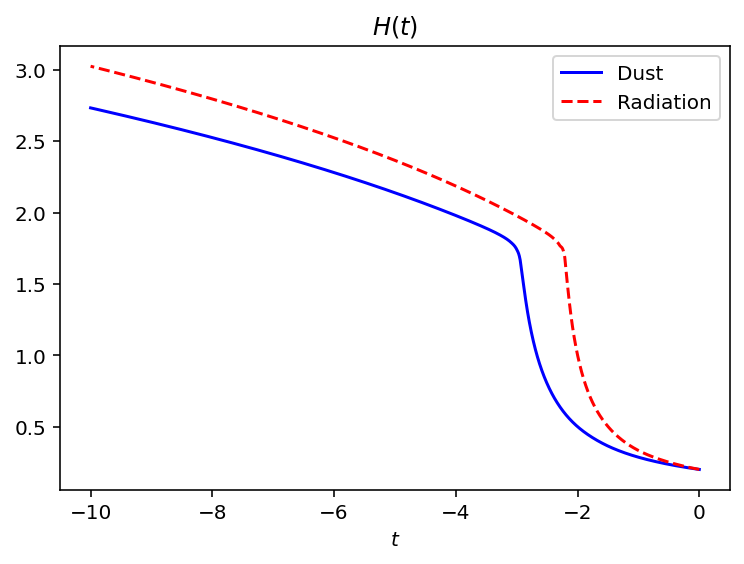} 
	\caption{Numerical integration of the acceleration equation (\ref{accele}) for Model 1 (\ref{model1}).}
	\label{fig:1}
\end{figure*}

The analysis of GI models usually focuses on radiation, $w=\frac{1}{3}$. In this work, however, dust, $w=0$, will take a prominent role since it is easier to handle when dealing with quantization. For comparison, we integrate numerically the acceleration equation (\ref{accele}) for Model 1 (\ref{model1}), with $k=0=\Lambda$, by fixing conditions at $t=0$ and evolving backwards in time. The results, shown in Fig. \ref{fig:1}, exhibit an inflationary stage exiting to the usual dust or radiation dominated eras. We are using testing values $(a_0=1,H_0=0.2)$ at $t=0$, $\la=10^{-6}$ and $L^{-1}=M_P=1$. For a discussion of realistic values of the parameters according to observational constraints see \cite{Edelstein_2020,JAIME2021100887}. 

For a barotropic fluid with linear equation of state, $p=w \rho$, the solution of continuity equation (\ref{continuity}) can be written as   
\begin{align}\label{sol1}
	\rho(t)=\rho_0\, \Big(\frac{a_0}{a(t)}\Big)^{3(1+w)}, 
\end{align}
where ($a_0,H_0,\rho_0$) are the variables evaluated at some reference time $t_0$. By the Friedmann constraint (\ref{frid}), only two initial conditions are independent, since $\rho_0=(3 F(h_0)-\Lambda)/\kappa^2$, with $h_0\equiv h(a_0,a_0 H_0)$. 

\begin{figure*}[ht]
	\centering
	\includegraphics[width=.4\textwidth]{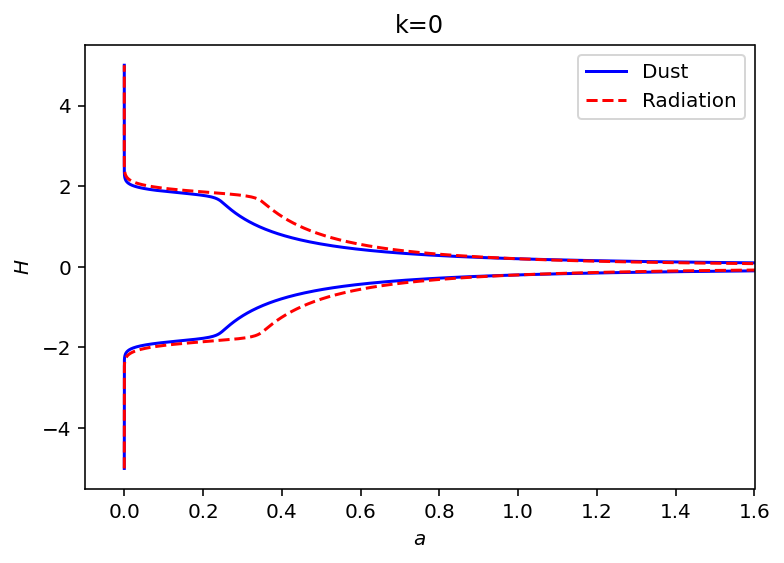}
	\qquad
	\includegraphics[width=.4\textwidth]{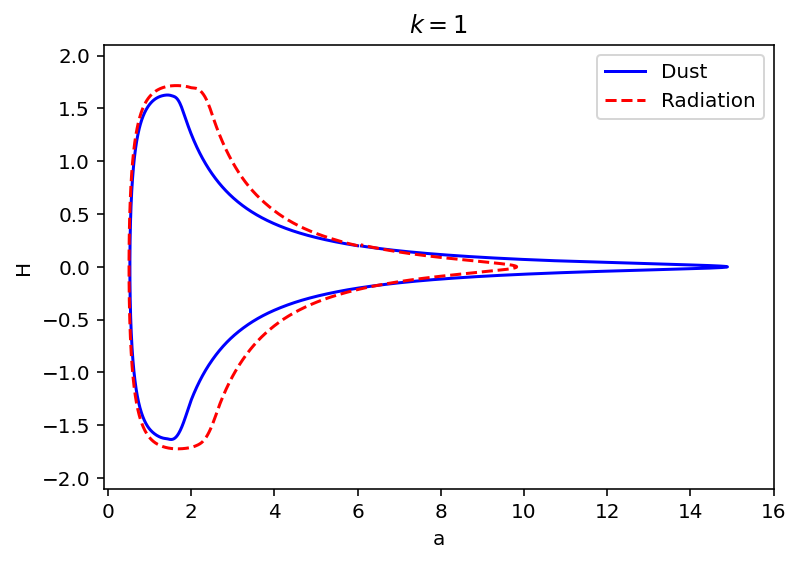}
	\caption{Phase curves for Model 1 (\ref{model1}) determined by the Friedmann constraint (\ref{frid}). For $k=0$, we choose them to pass through $(a_0, H_0)=(1,0.2)$. For $k=1$ they pass through $(a_0, H_0)=(6.0,0.2)$.}
	\label{averaver}
\end{figure*}

For the spatially flat case ($k=0$) and vanishing $\Lambda$, substituting (\ref{sol1}) into the Friedmann constraint (\ref{frid}) and solving for $a$, we obtain    
\begin{align}\label{solvea}
	a=a_0 \left(\frac{F(H_0)}{F(H)}\right)^{1/[3(1+w)]} \quad (w\ne -1)
\end{align}
In Fig. \ref{averaver}, we plot the curves in the plane $a-H$ determined by (\ref{solvea}) for Model 1 in the case of pressure free matter $w=0$ and radiation $w=\frac{1}{3}$. According to (\ref{solvea}), classical solutions do not cross the $H=0$ axis, the universe either expand or contract forever. For a contracting universe $H<0$, the time $\mathcal{T}$ to reach the singularity $a=0$, starting with finite $(a_0,H_0)$ at $t=t_0$, is given by 
\begin{align}\label{time}
	\mathcal{T}=\int_{a_0}^0 \frac{da}{a H}
	=\int_{-\infty}^{H_0} \frac{F'(H)/H}{3 (1+w) F(H)} dH
\end{align}
In the GR case the rightmost integrand is proportional to $2 H^{-2}$, thus we get a finite time. For models 1 and 2, however, the integrands tend, respectively, to $4 L^4 H^2$ and $2 L^2$ as $H\to \infty$, $\mathcal{T}$ is infinite.

On the other hand, for $k=1$, the Friedmann equation (\ref{frid}) reads, $F(h)=F(h_0) (a_0/a)^{3(1+w)}$. The fact that $h^2=H^2+a^{-2}$ makes it more difficult to solve for $a$ in terms of $H$. An alternative is drawing the vector field $(\at, \dot{H})$ tangent to the solution curves. This reveals that the phase curves do cross the axis $H=0$ at finite values $a$ given by the solutions of $a^{3+3 w}  F(\tfrac{1}{a}) =a_0^{3+3 w} F(h_0)$.

 In Fig. \ref{averaver}, we draw phase curves obtained with a numerical solution for $k=1$. Whereas with $k=0$ one can further into the past while approaching the classical singularity, for $k=1$, a minimum value of $a$ is reached where $H$ changes sign. In fact, we get a periodic motion between a minimum and a maximum value of scale factor (cf. \cite{Ling_2026} for a GQTG bounce model with $k=0$).



\section{Effective FRW action}\label{sec3}
Since we are interested in canonical quantization, the Hamiltonian formulation of FRW Geometric Inflation is in order. We construct here an effective 1D time-reparametrization invariant action for equations of motion (\ref{frid})-(\ref{accele}). For the matter component, we take the action principle for perfect fluids developed by Schutz \cite{PhysRevD.2.2762, PhysRevD.4.3559} and summarized in Appendix \ref{app1}.

FRW actions are usually obtained by evaluating the four-dimensional action at the FRW metric (\ref{frwmetric}) and integrating over spatial coordinates. In the case of geometric inflation, that would require to express the full action (\ref{generalized}) explicitly, however, only a finite number of terms $\mathcal{R}_n$ have been constructed \cite{ARCINIEGA2020135242}. Nonetheless, the FRW equations derived from these terms already exhibit the pattern (\ref{friedmann}), from which we can infer an FRW Lagrangian.

Since the Riemann tensor depends at most on second derivatives of the metric, we write
\begin{align}\label{action}
	L=\frac{3}{\kappa^2} N & \Big[a^2 D_t^2 a+a (D_t a)^2+k a-\frac{\Lambda}{3} a^3 \nonumber \\
 &+\sum_{n=3}^\infty \beta_n \tilde{L}_n \big(a, D_t a, D_t^2 a \big) \Big] 
\end{align}
where $\beta_n$ are constants and the covariant  derivatives (with respect to time transformations) of the scale factor are given by  
\begin{align}\label{covder}
	D_t a\equiv \frac{\at}{N}, && D_t^2 a\equiv \frac{\att^2}{N^2}-\frac{\at \Nt}{N^3}.
\end{align}

We use two requirements to fix $L_n\equiv N \tilde{L}$. First, the equation of motion of the scale factor must be second order in the gauge $N=1$ (we are excluding the case when the equation is second order only in a particular gauge $N\ne 1$ ). In consequence, the Lagrangian can be at most linear in acceleration \cite{einsteiniancubic}. Hence, we write 
\begin{align}\label{linear}
L_n=N  a^3 \Big[U_n(a, D_t a)+\frac{D_t^2 a}{a} V_n(a,D_t a)\Big].
\end{align}

Second, according to (\ref{frid})-(\ref{Fseries}), the contribution to the $k=0$ Friedmann equation from the $n$-th term of (\ref{action}) is proportional to $(\frac{D_t a}{a})^{2n}$.  Hence, varying (\ref{linear}) with respect to $N$ and rearranging, we obtain 
\begin{align}\label{deltaN}
U_n-D_t a \frac{\p U_n}{\p D_t a}+(D_t a)^2 \left(\frac{2 V_n}{a^2}+\frac{1}{a} \frac{\p V_n}{\p a}\right)=\left(\frac{D_t a}{a}\right)^{2n}.
\end{align} 
The latter equation can be satisfied in several ways. We make the following choice
\begin{align}
\begin{split}
&U_n=\big(\tfrac{D_t a}{a}\big)^{2n},\\
&V_n=y_n \big(\tfrac{D_t a}{a}\big)^{2n-2}, \quad y_n=\frac{n}{2-n}, \quad n\ne 2,
\end{split}
\end{align}
since it reproduces the expressions for $n=0, 1$ and $n=3$ \cite{einsteiniancubic}. 

To include spatial curvature, we only need to replace $\frac{D_t a}{a}$ in (\ref{linear}) with $h(a,D_t a)$ (see definition (\ref{hchikis})). Therefore, the $n$-th term of our FRW Lagrangian is 
\begin{align}\label{lnth}
	L_n =\beta_n N a^3 \left[h^{2 n}+y_n h^{2 n-2} \frac{D_t^2 a}{a} \right],
\end{align}
where $\beta_n\equiv (-)^n \la_n L^{2n-2}$. Substituting (\ref{lnth}) into (\ref{action}), and rearranging, we get  
\begin{align}\label{frwgeometric}
	L=\frac{3}{\kappa^2} N a^3 \left.  \Big[F(h)+\frac{D_t^2 a}{a} h^2 J(h)-\frac{1}{3} \Lambda\Big]\right|_{h=h(a,D_t a)},
\end{align}
where $F(h)$ is defined in (\ref{Fseries}) and
\begin{align}
	J(h)\equiv h^{-2}+\sum_{n=3}^\infty y_n \beta_n h^{2 n-4} =-\int dh \frac{F'(h)}{h^4}
\end{align}

It can be readily shown that Lagrangian (\ref{frwgeometric}) yields the gravitational part of the Friedmann equations (\ref{frid})-(\ref{accele}). 

Let's set $N=1$ for a moment so that $D_t a=\at$ and $D_t^2 a=\att$. Further, performing the variation of the action (as a higher-derivative system) yields boundary terms proportional to $\delta \at$ only, not to $\delta a$. For a variational principle with fixed $a$ on the boundary, we redefine the problem as follows. First, we write the term of (\ref{frwgeometric}) depending on $\att$ as $\frac{\p g}{\p \at}$, for a function $g(a, \at)$. Next, define a new Lagrangian $\tilde{L}=L-\at \frac{\p G}{\p a}-\att \frac{\p G}{\p \at}=L-\at \frac{\p G}{\p a}$, that no longer depends on acceleration. A quick look at (\ref{frwgeometric}) tell us that, up to a function of $a$, 
\begin{align}
	G(a,\at)&=\frac{3}{\kappa^2} \int d\at \big[ \big(\at^2+k\big) J(h)|_{h(a,\at)} \big] 
\end{align}
Thus, $\frac{\p G}{\p a}\propto \int d\at \, (\at^2+k) \frac{dJ}{dh} \frac{\p h}{\p a}=\int d\at  a \frac{F'(h)}{h}$, and 
\begin{align}\label{alter}
	\tilde{L}
	&=\frac{3}{\kappa^2} a^3 \Big[F(h)-\frac{\Lambda}{3}-\frac{\at}{a^2} \int d\at \left. \frac{F'(h)}{h}\right|_{h(a,\at)}\Big].
\end{align}
For the covariant expression one simply replaces $\at$ with $D_t a$, and the global factor $a^3$ with $N a^3$.

\subsection{Hamiltonian formulation}
Lagrangian (\ref{alter}) allows one to perform the usual Hamiltonian formulation. However, the momentum conjugate to the scale factor, namely,  
\begin{align}\label{tildepa}
	\tilde{p}_a\equiv \frac{\p \tilde{L}}{\p \at} 
	=-\frac{3 a}{\kappa^2} \left. \left(\int ds  \frac{F'(h(a,s))}{h(a,s)}\right)\right|_{s=\at/N},
\end{align}
is a complicated function of $(a,\at)$, even with the simplest cubic model \cite{einsteiniancubic}. Instead, we start with the higher-derivative Lagrangian (\ref{frwgeometric}), whose explicit dependence on acceleration fits to Ostrogradsky's construction \cite{Woodard2007}. The Lagrange multiplier method, commonly used with $F(R)$ theories, would yield an equivalent canonical formulation.

In Ostrogradsky's approach, for every second time derivative, there is an additional canonical pair. Since (\ref{frwgeometric}) does not depend on $\ddot{N}$, we have three canonical pairs with coordinates $N, a, v\equiv \at$, and conjugate momenta 
\begin{subequations}\label{ostrmomenta}
\begin{align}
	p_N&\equiv \frac{\p L}{\p \Nt}=-\frac{3}{\kappa^2} \frac{a^2 \at h^2 J(h)}{N^2}, \\
	p_a&\equiv \frac{\p L}{\p \at}-\frac{d}{dt} \frac{\p L}{\p \att}=0, \label{pa}\\
	p_v&\equiv \frac{\p L}{\p \att}=\frac{3}{\kappa^2} \frac{a^2 h^2 J(h)}{N}.  \label{pv}
\end{align}
\end{subequations}
with $h=h(a,\nu/N)$. The canonical Hamiltonian is
\begin{align}
	H&\equiv \Nt p_N+\at p_a+v p_v-L \nonumber \\
	&=\frac{3}{\kappa^2} N a^3 \left[\frac{\Lambda}{3}-\frac{\frac{\nu^2}{N^2}+k}{a^2}-\sum_{n=3}^\infty
	\beta _n \frac{\big(\frac{\nu^2}{N^2}+k\big)^n}{a^{2n}}\right].
\end{align}
Recall $\beta_n\equiv (-)^n \la_n L^{2n-2}$.

At this point, it is convenient to perform the following canonical transformation 
\begin{align}\label{newcanonical}
\left(N, \tilde{p}_N\equiv p_N+\frac{v}{N} p_v\right), &&  \left(B\equiv \frac{v}{N}, p_B\equiv N p_v\right), 
\end{align}
and ($a,p_a$) stays the same. Then $H=N H_0(a,N)$, with 
\begin{align}\label{h0}
	H_0 (a,B)
	=\frac{a^3}{\kappa^2} [\Lambda-3 F(h)] 
\end{align}
where, from now on, $h=h(a,B)=\frac{\sqrt{B^2+k}}{a}$. 

From definitions (\ref{ostrmomenta}) we get three primary constraints, 
\begin{subequations}
\begin{align}
	C_1 &\equiv \tilde{p}_N \approx 0, \\
	C_2 &\equiv p_a \approx 0, \\
	C_3 
	&\equiv p_B-\frac{3}{\kappa^2} a^2 h^2 J(h)\approx 0
\end{align}
\end{subequations}

The total Hamiltonian is $H^{\text{tot}}=N H_0+u_1 C_1+u_2 C_2+u_3 C_3$, with multipliers $u_i$ that we try to determine using the consistency conditions \cite{henneaux}: 
\begin{subequations}
\begin{align}
	\dot{C}_1&=\{C_1,H^{\text{tot}}\}=-H_0(a,B) \equiv C_4 \approx 0, \label{consis1} \\
	\dot{C}_2&=\{C_2,H^{\text{tot}}\}=-N \frac{\p H_0}{\p a}-u_3 \frac{\p C_3}{\p a} \approx 0, \label{consis2} \\ 
	\dot{C}_3&=\{C_3,H^{\text{tot}}\}
	=-N \frac{\p H_0}{\p B}+\frac{\p C_3}{\p a} u_2 \approx 0. \label{consis3}
\end{align}
\end{subequations}
Thus, (\ref{consis1}) yields another constraint $C_4$, (\ref{consis2}) yields $u_3=N a \frac{\Lambda-3 F(h)}{F'(h)/h}+N a h^2$ and (\ref{consis3}) yields $u_2=B N$. 

We classify the constraints into first class, $\phi_1=N H_0+u_2 p_a+u_3 C_3$ and $\phi_2=C_1$, and second-class, $\chi_1=C_2$ and $\chi_2=C_3$. Counting degrees of freedom yields $n_{\text{dof}}=3-2 (\text{first-class})-\frac{2 (\text{second-class})}{2}=0$, as with the ordinary FRW model.

We get rid of $p_a$ and $p_B$ by solving the second-class constraints and using the basic Dirac brackets $\{N,\tilde{p}_N\}_D=1$ and  
\begin{align}\label{aBF}
	\{a,B\}_D=-\frac{\kappa^2 }{3 a f(h)}, 
\end{align}
where $f(h)$ is the following non-vanishing function 
\begin{align}\label{fdeh}
	f(h)\equiv \frac{F'(h)}{h}=2+ \sum_{n=3}^\infty 2n \beta_n h^{2n-2}.
\end{align}

\section{Quantum geometric inflation}\label{sec4}
In this section, we address the Wheeler-DeWitt quantization of the spatially flat geometric inflation models (\ref{model1})-(\ref{model2}) based on the canonical formulation just described. For the perfect fluid part we use Schutz's formalism, summarized in Appendix \ref{app1}. The reader is referred to \cite{PhysRevD.2.2762,PhysRevD.4.3559,lemos,Lapchinskii1977} for more details.

Given the nontrivial Dirac bracket (\ref{aBF}), we look for an actual canonically conjugated pair $(A,P)$, which allows us to use the standard position or momentum representations. For $k=0$, the basic Dirac bracket (\ref{aBF}) reads $\{a,B\}_D=-\kappa^2/[3 a f(B/a)]$, it can be readily shown that the following pair,   
\begin{align}\label{AP}
	A(a,B)=\frac{a^3}{\kappa^2} f\big(\tfrac{B}{a}\big), && P(a,B)=-\frac{B}{a}, 
\end{align}
is canonically conjugated, that is, $\{A,P\}_D=1$. The minus sign in the new momentum is chosen so that $A$ is non-negative. 

In terms of the new  variables (\ref{AP}), the spatially flat pure gravity Hamiltonian constraint (\ref{h0}) becomes, 
\begin{align}\label{fullhamil1}
	H_0^{\text{grav}}=A \, \frac{\Lambda-3 F(P)}{f(P)}  \approx 0,
\end{align}
where we are taking even $F(P)$ (hence even $f(P)$).

In the following we shall set $\Lambda=0$ and consider instead a perfect fluid with a barotropic linear equation of state $p=w \rho$. The perfect fluid contribution to the Hamiltonian constraint can be read off from (\ref{hfluid}). The full Hamiltonian constraint is therefore   
\begin{align}\label{fullhamil2}
	H_0=-3 A \frac{F(P)}{f(P)}+\left(\frac{f(P)}{\kappa^2 A}\right)^w  P_T \approx 0 
\end{align}
where $(T,P_T)$ is the canonical pair representing the only relevant fluid's degree of freedom. 

\subsection{Dust: Position representation}\label{wkb}
Setting $w=0$, (\ref{fullhamil2}) reduces to $H_0=-3 A F(P)/f(P)+P_T$. In the position representation $\hat{P}  \Psi(A)=-i \hbar \p_A \Psi(A)$ and $f^{-1}(P)$ is promoted to a non-local inverse operator \cite{Teschl}. To avoid this complication, instead of quantizing directly with $H_0$, we use another quantity proportional to it, namely,  
\begin{align}\label{newconstraint}
	\mathcal{H}\equiv -3 F(P)+A^{-1} f(P) P_T \approx 0.
\end{align}

Now, following Dirac quantization scheme, we postulate the existence of a wave function  $\Psi(T,A)$ to be annihilated by the Hamiltonian constraint operator or, in this case, $\hat{\mathcal{H}} \Psi=0$, which is our Wheeler-DeWitt equation. With symmetric ordering, it reads 
\begin{align}\label{tisedust}
3 F(\hat{P}) \Psi(T,A)=-i \hbar	\big[A^{-1} f(\hat{P})+f(\hat{P}) A^{-1}\big] \frac{\p \Psi(T,A)}{ \p T}.
\end{align}
Substituting a separable wave function  
\begin{align}\label{separaA}
	\Psi(T,A)=\exp \left(\tfrac{i}{\hbar} \chi \, T\right)  U(A), 
\end{align}
and simplifying, we obtain the following $T$-independent equation 
\begin{align}\label{tise}
	3 F(\hat{P}) U(A)=\frac{\chi}{2} \left[A^{-1} f(\hat{P})+f(\hat{P}) A^{-1}\right] U(A) 
\end{align}
where $\chi$ is the separation constant. 

In the WKB approach below, the zeroth order Hamilton-Jacobi equation (\ref{zeroth}) yields $\chi>0$. Thus, states with classical counterparts have positive $\chi$.

For GR, $F(P)=P^2$ and $f(P)=2$. Thus, (\ref{tise}) yields a second order differential equation, 
\begin{align}
\p_A^2 U(A)+\frac{2 \chi}{3 \hbar^2 A} U(A)=0
\end{align}
whose general solution involves Bessel functions of the first and second kinds, $U=c_1 y J_1(2y)+c_2 y Y_1(2y)$, where $y=\sqrt{2 \chi A/(3 \hbar^2)}$ (cf. \cite{lemos}).

With GI models such as (\ref{model1}), $F(P)$ and $f(P)$ are nonpolynomial functions and we obtain an infinite order differential equation. Thus, we set up a WKB expansion \cite{liboff}, for which a finite order equation can be solved at every order of the WKB approximation. Alternatively, we may consider the asymptotic behavior of the function classical quantity $F(P)/f(P)$ for large and small $P$. This is appropriate in the momentum representation, addressed in Section \ref{sec5}. 

Substituting $U(A)=e^{i S(A)/\hbar}$, with $S(A)=S_0(A)+\hbar S_1(A)+\mathcal{O}(\hbar^2)$ into (\ref{tise}), and using the series (\ref{Fseries}) for  $F(\hat{P})$, we obtain  
\begin{align}\label{expansion}
	F(\hat{P}) U(A)=&\Big[ F(S'_0)+\hbar \Big(S'_1(A) F'(S'_0)  \nonumber \\
	& -\tfrac{i}{2} S''_0(A) F''(S'_0)\Big)+\mathcal{O}(\hbar^2)\Big]  U(A),
\end{align}
where $S_0'\equiv \frac{dS_0}{dA}$, $S_1'\equiv \frac{dS_1}{dA}$ and $F'(S_0'), F''(S_0')$ stand for the first and second derivatives of $F$ with respect to its argument. An expression analogous to (\ref{expansion}) is obtained for $f(\hat{P}) U(A)$. On the other hand, for $f(\hat{P}) (A^{-1} U)$, we obtain    
\begin{align}
	f(\hat{P}) A^{-1} U(A)&=\left\lbrace \frac{f(S'_0)}{A}+\hbar \Big[\frac{S'_1(A) f'(S'_0)}{A}+i \frac{f'(S'_0)}{A^2} \right.  \nonumber \\
	& \left. -\frac{i}{2} \frac{S''_0(A) f''(S'_0)}{A} \Big]+\mathcal{O}(\hbar^2) \right\rbrace U(A)
\end{align}

In this way, we find, to zeroth order in $\hbar$, the $T$-independent Hamilton-Jacobi (H-J) equation 
\begin{align}\label{zeroth}
	F(S_0')=\frac{\chi}{3 A} f(S_0')
\end{align}

On the other hand, to first order in $\hbar$, we get
\begin{align}
	S'_1(A)=\frac{i}{2} \frac{S''_0 \left( F''(S'_0)-\frac{\chi}{3 A} f''(S'_0)\right) +\frac{\chi}{3 A^2} f'(S'_0)}{F'(S'_0)-\frac{\chi}{3 A} f'(S'_0)}
\end{align}
Thus, up to an additive constant, 
\begin{align}\label{sorder1}
	S_1(A)=\frac{i}{2} \ln \left| F'(S'_0)-\frac{\chi}{3 A} f'(S'_0)\right| 
\end{align}

Therefore, the first order WKB approximation is 
\begin{align}\label{wkbfirst}
	U(A)
	&=\frac{C \exp (\frac{i}{\hbar} S_0)}{\sqrt{|F'(S'_0)-\frac{\chi}{3 A} f'(S'_0)|}}
\end{align}

Further, using the derivative of (\ref{zeroth}) with respect to $A$ and assuming $S_0''\ne 0$,   
\begin{align}\label{wkbgen}
	U(A)=\left| \frac{A S_0''(A)}{F(S_0')}\right|^{1/2}  \exp \left[ \frac{i S_0(A)}{\hbar} \right] 
\end{align}

For Model 1 (\ref{model1}), the H-J equation (\ref{zeroth}) can be written as 
\begin{align}\label{hjmod0}
	A=\frac{\chi}{3} \frac{2+4 \la (2 L^6 P^6+L^{10} P^{10}) e^{L^4 P^4}}{P^2+\la L^6 P^8 e^{L^4 P^4}}
\end{align}
where $P=S_0'(A)$ and $\la>0$.

\begin{figure}[ht]
	\centering
	\includegraphics[width=.42\textwidth]{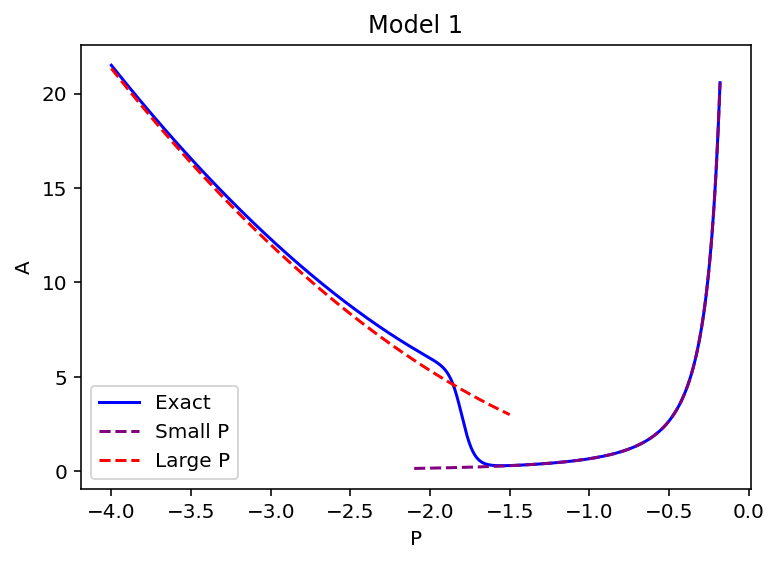}
	\caption{$A(P)$ as obtained from (\ref{hjmod0}) with $\la=10^{-6}$, $\chi=1.0$. Note that $A$ is bounded from below for $\la\ne 0$.}
	\label{fig:18c}
\end{figure}

Fig. \ref{fig:18c} shows the right hand of (\ref{hjmod0}), together with their asymptotic behavior for large and small $|P|$,
\begin{align}\label{aapprox}
	A \approx 
	\begin{cases}
		\frac{2 \chi}{3 P^2}, & |P|\ll 1, \\
		\frac{4}{3} \chi L^4 P^2, & |P|\gg 1
	\end{cases}
\end{align}

Thus we approximate the two independent real roots, $P=S_0'(A)$, of (\ref{hjmod0}) as follows,  
\begin{subequations}\label{s0approx}
	\begin{align}
		[S'_0(A)]_1&\approx \pm \sqrt{\frac{2 \chi}{3 A}}, \label{s1} \\
		[S'_0(A)]_2& \approx \pm \frac{\sqrt{3 A}}{2 L^2 \sqrt{\chi}},\label{example1}
	\end{align}
\end{subequations}
both in the regime of large $A$. 

Since $S_0'' \ne 0$ for (\ref{s0approx}), using (\ref{wkbgen}) with the corresponding small/large $P$ approximations of $F(P)$, we get four independent WKB wave functions
\begin{subequations}\label{usss}
	\begin{align}
		U_1(A)&\approx \left(\frac{3 A}{8\chi}\right)^{1/4}   \exp \left[\pm \frac{2 i}{\hbar} \sqrt{\frac{2 \chi}{3}} A^{1/2}\right], \\
		U_2(A)&\approx \frac{8 L^4}{\sqrt{\lambda }} \left(\frac{\chi }{3 A}\right)^{7/4} \exp \left[-\frac{9 A^2}{32 L^4 \chi^2}\pm \frac{i}{\hbar} \frac{A^{3/2}}{\sqrt{3\chi} L^2} \right], \label{u2}
	\end{align}
\end{subequations}
valid in the region $A>A_{\text{min}}$. The extra exponential term in (\ref{u2}), coming from the quantum prefactor (\ref{sorder1}), suppresses the amplitude for large $A$. 

\begin{figure}[ht]
	\centering
	\includegraphics[width=.44\textwidth]{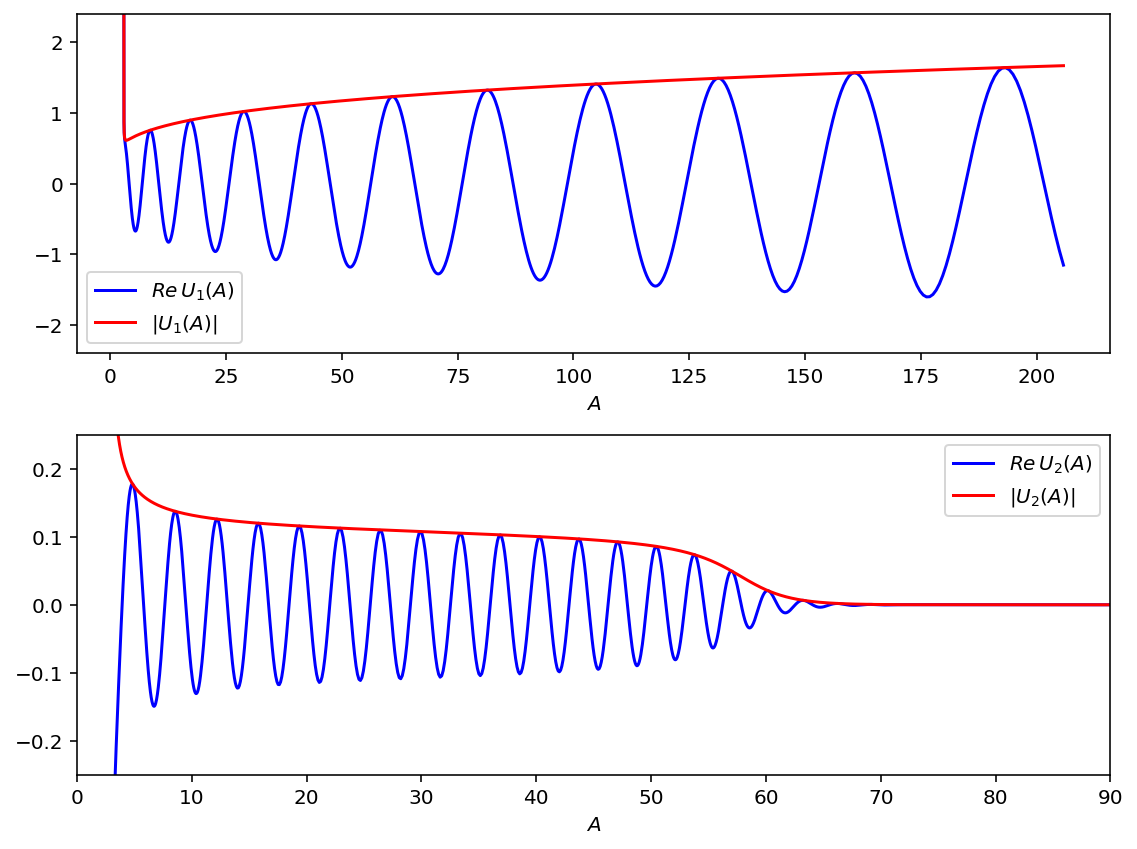}
	\caption{Two independent WKB solutions. }\label{fig:24}
\end{figure}

As can be seen in Fig. \ref{fig:18c}, large $|P|$ approximation starts failing as $A$ reaches it minimum while $P$ is approximately constant. Thus, we expect that, as $A$ gets closer to its classical minimum value, $U_2$ resembles a plane wave. We can verify this by integrating numerically, $S_0(A)=\int_{A_0}^A S_0'(A') dA'=\int_{A_0}^A P(A') dA'$, with one integral for each single-valued branch of Fig. \ref{fig:18c}. The resulting WKB wave functions obtained in this way are shown in Fig. \ref{fig:24}.

Quantum theory allows nonzero amplitude for $A\le A_{\text{min}}$. For ordinary system, a classical turning point occurs at vanishing momentum, and the analytic continuation proceeds with imaginary momentum \cite{liboff,einsteiniancubic}. Here the minimum is reached a finite $P$ (Fig. \ref{fig:18c}) and we should consider complex $P$. We do not pursue this further and instead, work directly in momentum space, where the high and small curvature regimes are are described by single wave function instead of the two branches of Fig. \ref{fig:24}.

The phase of the WKB wave functions (\ref{usss}) contains information about the classical dynamics described in Section \ref{sec2}. For inflation, we take the large momentum approximation (\ref{u2}), so that $\Psi_2(T,A)\propto \exp \frac{i}{\hbar} \big[\chi T\pm \frac{A^{3/2}}{\sqrt{3\chi} L^2}\big]$. First, we differentiate the phase with respect to $\chi$ and set the result equal to a constant, $T \pm \frac{1}{2 \sqrt{3} L^2} (\frac{A}{\chi})^{3/2}=T_1$. Then, using the linear relation between $T$ and $t$ (see \ref{derit}), $|T_1-T|=|t_0-t|$ and $A(t)=(2 \sqrt{3} L^2)^{2/3} \chi |t_0-t|^{2/3}$. Next, according to (\ref{AP}), in the gauge $N=1$, $H=-P=-\frac{\p S_0}{´\p A}$, that is, 
\begin{align}\label{solclass}
	H=\mp \frac{1}{2 L^2} \left(\frac{3 A}{\chi}\right)^{1/2}=\mp \left(\frac{3}{2 L^2}\right)^{2/3} |t_0-t|^{1/3},
\end{align}
with $t_0>t$. Thus, (\ref{solclass}) satisfies $\dot{H}=-3/(4L^4 H^2)$, which is the large $H$ asymptotic form of acceleration equation (\ref{accele}) for Model 1 and $a(t)=\exp[-\frac{3}{4} \big(3/(2 L^2)\big)^{2/3} (t_0-t)^{4/3}]$, which yields a curve $a(t)$ resembling that of Fig. \ref{fig:1}.

\section{Momentum representation}\label{sec5}
In this section, we compute the wave functions directly in the momentum representation using  $\hat{A} \Phi(P)=i \hbar \p_P \tilde{\Psi}(P)$ and $\hat{P} \tilde{\Psi}(P)=P \Phi(p)$ (for the fluid variables we still use $\hat{P}_T=-i \hbar \p_T$). Unlike section (\ref{wkb}), the original Hamiltonian constraint (\ref{fullhamil2}), with $w=0$, is well enough to work with, 
\begin{align}\label{hamilP}
H_0&=-A G(P)+P_T \approx 0.
\end{align}
where we defined    
\begin{align}\label{sch1}
	G(P)\equiv \frac{3 F(P)}{f(P)}=\frac{3 F(P)}{F'(P)/P}
\end{align}

The $T$-dependent WDW equation, $\hat{H}_0 \Phi(P)$, thus reads 
\begin{align}\label{wdw}
i \hbar \frac{\p \tilde{\Psi}(T,P)}{\p T}=-\frac{\hat{A} G(P)+G(P) \hat{A}}{2} \tilde{\Psi}(T,P)
\end{align}
Substituting the separable wave function 
\begin{align}\label{psitp}
	\tilde{\Psi}(T,P)=\exp (\tfrac{i}{\hbar} \xi T) \Phi(P),
\end{align}
and simplifying, we obtain the following first order differential equation,
\begin{align}
 \frac{1}{2} G'(P) \Phi(P)+G(P) \Phi'(P)=\frac{\xi}{i \hbar} \Phi(P),
\end{align}
whose formal solution is given by 
\begin{align}\label{momenwave}
	\Phi(P)=\frac{C}{\sqrt{G(P)}}\exp \left[ \frac{i \xi}{\hbar} \left(-\int \frac{dP}{G(P)} \right)\right]. 
\end{align}
with $C$ a complex constant. The sign of separation constant $\xi$ is discussed below. 

The wave function's modulus is thus determined by $G(P)$ (\ref{sch1}) and the remaining problem is to perform the integral (\ref{momenwave}) to get the phase. We make first some approximations to guide our subsequent numerical computations. Since the classical variable $P(t)=-H(t)$ does not change sign for $k=0$ (see Fig. \ref{averaver}), we only consider solutions for negative $P$, or equivalently, positive $H$. 

For Model 1 (\ref{model1}) we have  
\begin{align}\label{w1m1}
	G^{-1}(P)=\frac{2+4 \la (2 L^6 P^6+L^{10} P^{10}) e^{L^4 P^4}}{3 (P^2+\la L^6 P^8 e^{L^4 P^4})}
\end{align}
which is an even function of $P$ (see Fig. \ref{fig:18c}), diverging at $P=0$ and at $|P|\to \infty$. Since the classical solutions do not cross the $P=-H=0$ axis, it is fair to restrict our domain to the negative $P$ semi-axis. We may use the asymptotic forms already used in  (\ref{aapprox}), namely 
\begin{align}\label{gm1app}
	G^{-1}\approx 
	\begin{cases}
		\frac{4}{3} L^4 P^2, \quad &|P|\gg 1 \\
		\frac{2}{3 P^2}, \quad &|P|\ll 1
	\end{cases}
\end{align}
The integral of (\ref{momenwave}) can easily be performed to yield the approximate wave function, for $P<0$, 
\begin{align}\label{wkbmod1}
	\Phi(P) \approx 
	\begin{cases}
		\frac{-2 L^2 P}{\sqrt{3}} \exp \big[\frac{i \xi}{\hbar} \big(-\frac{4}{9} L^4 P^3\big) \big], \quad &|P|\gg 1, \\
		-\frac{\sqrt{2/3}}{P} \exp \big[\frac{i \xi}{\hbar} \frac{2}{3 P}\big], \quad \quad &|P|\ll 1
	\end{cases} 
\end{align}

\begin{figure}[ht]
	\centering
	\includegraphics[width=.4\textwidth]{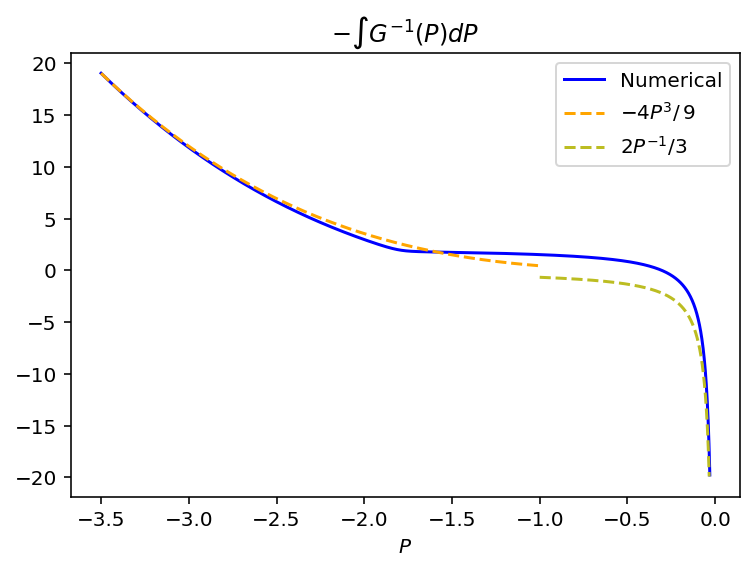}
	\caption{Numerical integration of the phase (\ref{momenwave}) for Model 1.}
	\label{fig:34a}
\end{figure}

Now, for the transition from large to small $P$, we perform a numerical evaluation of the integral $I(P) \equiv \int_{P_0}^P (-G^{-1} (P') dP')+c_1$. The constant where $c_1$ is chosen so that at the initial point $P_0$, $I(P)$ coincides with the large momentum approximation, $I(P_0)=-\frac{4 L^4}{9} P_0^3$. A comparison is shown in Fig. \ref{fig:34a}. 

\begin{figure}[t]
	\centering
	\includegraphics[width=0.4\textwidth]{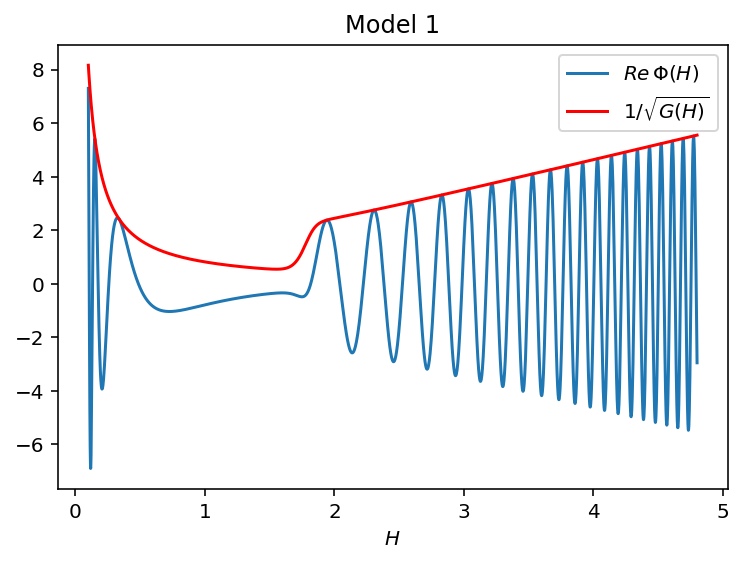}
	\hfill
	\includegraphics[width=0.4\textwidth]{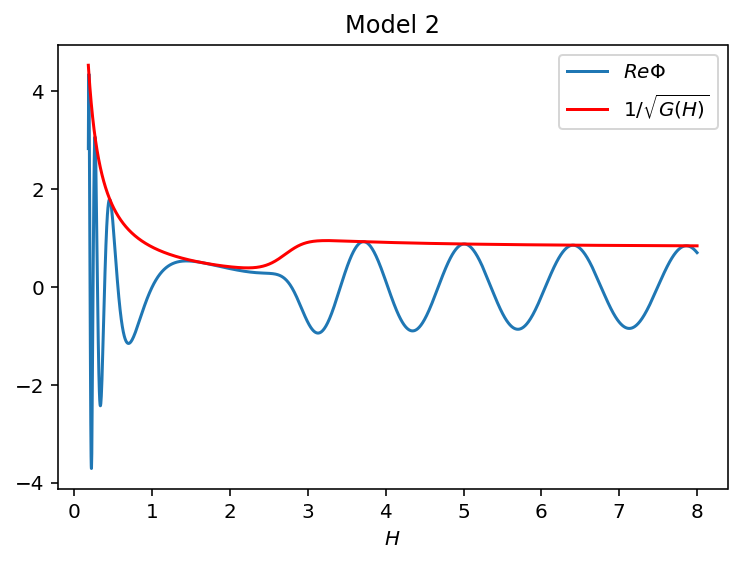}
	\caption{Momentum space wave functions. We use $\xi=2.6$ for Model 1 and $\xi=6$ for Model 2.}
	\label{fig:36}
\end{figure}

Since the Hubble factor $H$ is a more familiar quantity, after obtaining $\Phi(P)$, we switch to $H$, $\tilde{\Phi}(H)=(G(H))^{-1/2} \exp[\frac{i}{\hbar} \xi I(-H)]$ and plot the result in Fig. \ref{fig:36}.  The infinite oscillations over $0<H<1$ can be better appreciated plotting against $\ln H$.

On the other hand, for Model 2 (\ref{model2}), we have 
\begin{align}\label{gm2app}
	G^{-1}=\frac{2 \big(1-\lambda [3 L^4 P^4-L^6 P^6 (4+L^2 P^2) e^{L^2 P^2}]\big)}{3 \big(P^2-\lambda [L^4 P^6-L^6 P^8 e^{L^2 P^2}]\big)}
\end{align}
The small $P$ regime is the same as in (\ref{wkbmod1}), whereas for large $|P|$, $G^{-1}\approx \frac{2}{3} L^2$, a constant, then
\begin{align}\label{wkbmod2}
	\Phi(P)\approx  \sqrt{\frac{2}{3}} L \exp \left[ i \frac{\xi}{\hbar} \left(-\frac{2}{3} L^2 P\right)\right], \quad |P|\gg 1
\end{align}
The numerical integration of the phase is computed as for Model 1. The corresponding wave function is shown in Fig. \ref{fig:36} and, as expected, it resembles a plane wave for large $H$.

Before putting these momentum space wave functions to work, let's consider the semi-classical expectations. We can obtain the classical dynamics using a mixed H-J formalism, $H_0(A=-\p_P \mathcal{S}, P, T, P_T=\p_T \mathcal{S})=0$, with Hamiltonian (\ref{hamilP}). Setting up  $\mathcal{S}(P,T,\xi)=\xi T-V(P)$ yields $A=-V'(P)=\frac{\xi}{G(P)}$. From which we learn that $\xi>0$ and $\mathcal{S}=\xi T-\xi \int G^{-1}(P) dP$ is the phase of wave functions (\ref{psitp}). Using the large momentum approximation of (\ref{gm1app}), we have $\mathcal{S}\approx \xi (T-\frac{4}{9} L^4 P^3)$. Now, setting $\p_\xi \mathcal{S}=T_1$ yields $\frac{4 L^4}{9}  P^3 \equiv -(T_1-T)$ 
$P=-H \propto -(T_1-T)^{1/3}$, in agreement with (\ref{solclass}).

\subsection{T-dependent wavepacket}
Since the individual momentum wave functions (\ref{momenwave}) are not normalizable in $H \in (0, \infty)$, we expect that a suitable linear combination has sufficient interference to cancel the divergent part occurring at small and large $H$.

Let's construct Gaussian wave packets  with the  momentum space $\xi$-dependent wave functions (\ref{momenwave}) as follows,   
\begin{align}\label{wavepacket}
\Psi(P,T) &=\int_0^\infty \mu(\xi) \tilde{\Psi}(T,P,\xi)\, d\xi, 
\end{align}
where the Gaussian distribution $\mu(\xi)$ has variance $\sigma$ and is centered at $\bar{\xi}$,  that is, 
\begin{align}\label{distro}
	\mu(\xi)=\frac{\exp[-\frac{1}{2 \sigma^2} (\xi-\bar{\xi})^2]}{\sqrt{2\pi} \sigma} 
\end{align}

We proceed with some analytic approximations and complement them with numerical computations. For Model 1, using (\ref{psitp}) and the large $P$ approximation (\ref{wkbmod1}), we have  $\tilde{\Psi}(T,P,\xi)\approx -\frac{2}{\sqrt{3}} L^2 P \exp [\frac{i \xi}{\hbar} (T-\frac{4}{9} L^4 P^3)] $ or, in terms of $H$,
\begin{align}\label{solp1}
	\tilde{\Psi}(T,H)\approx \frac{2}{\sqrt{3}} L^2 H \exp \Big[\frac{i \xi}{\hbar} \big(T+\frac{4}{9} L^4 H^3\big)\Big] 
\end{align}
Next, we perform the following Gaussian integral\footnote{Note that we integrate over the full $\xi$-axis. Integrating over $[0,\infty)$ gives the complex exponential in (\ref{wave3d1}) with a factor $\frac{1}{2}[1+\text{erf} (x+i y)]$, where $x=\bar{\xi}/(\sqrt{2}\sigma)$ and $y=\sigma (T+\frac{4}{9} L^4 H^3)/\sqrt{2}$. However, for large $x$ and finite $y$, the error function has a series expansion $1-\frac{1}{\sqrt{\pi} x} e^{-(x+i y)^2} (1+\mathcal{O}(\frac{1}{x}))$. In this scenario, setting the extra factor equal is a convenient approximation.}, 
\begin{align}\label{wave3d1}
	\Psi(T,H)& \equiv \int_{-\infty}^\infty \mu(\xi) \tilde{\Psi}(T,H) d\xi \nonumber \\ 
	&=\frac{2 L^2 H}{\sqrt{3}} e^{-\frac{\sigma^2}{2} (T+\frac{4}{9} L^4 H^3)^2+ i \bar{\xi} (T+\frac{4}{9} L^4 H^3)}.
\end{align}

The WDW equation (\ref{wdw}) lends the fluid variable $T$ the character of an evolution parameter for $\hat{H}$ \cite{lemos,Burton-Villalobos2025}. Thus, in the following we treat $T$ as an external parameter and define a normalized $T$-dependent wave function
\begin{align}\label{newwave}
	\psi(T,H)\equiv \frac{\Psi(T,H)}{\sqrt{\int_0^\infty \Psi^*(T,H) \Psi(T,H) dH}}\equiv \frac{\Psi(T,H)}{\mathcal{N}(T)}
\end{align}

For example (\ref{wave3d1}), we get
\begin{align}\label{normat}
\mathcal{N}^2(T)
&=\frac{4 L^4}{3} \int_0^\infty H^2 e^{-\sigma^2 (T+\frac{4}{9} L^4 H^3)^2} dH \nonumber \\
&=\frac{\sqrt{\pi}}{2 \sigma} \text{erfc}(\sigma T) 
\end{align}

Next, we compute the mean value of $\hat{H}$ using the normalized wave function (\ref{newwave})
\begin{align}\label{meandef}
\braket{\hat{H}}(T)&\equiv \int_0^\infty \psi^*(T,H)\, H \, \psi(T,H) dH \nonumber \\ 
&=\frac{4 L^4}{3 \mathcal{N}^2(T)} \int_0^\infty H^3 e^{-\sigma^2 (T+\frac{4}{9} L^4 H^3)^2} dH.
\end{align}
Making the change of variable $u=\frac{4}{9} \sigma L^4 H^3$, we get 
\begin{align}\label{specialinte}
	\braket{H}(T)&=\frac{(\frac{3}{2})^{2/3} e^{-\sigma^2 T^2}}{(L \sigma)^{4/3} \mathcal{N}^2(T)} \int_0^\infty u^{1/3}  e^{-(u^2+2u \sigma T)} du
\end{align}
The last integral is given by a Kummer confluent hypergeometric function (cf. Appendix \ref{app2}), 
\begin{align}\label{meanapp}
\braket{H}(T)&=\frac{(\frac{3}{2})^{2/3} \exp(-\sigma^2 T^2)}{L^{4/3} \sigma^{1/3} \sqrt{\pi} \, \text{erfc}(\sigma T)} \left[\Gamma(\tfrac{2}{3}) \, _1 F_1(\tfrac{2}{3}; \tfrac{1}{2}, \sigma^2 T^2) \right. \nonumber \\
& \quad \quad \quad \left. -2 \, \sigma T \, \, \Gamma(\tfrac{7}{6}) \, _1 F_1(\tfrac{7}{6}; \tfrac{3}{2}, \sigma^2 T^2) \right] 
\end{align}

\begin{figure*}[ht]
	\centering
	\begin{subfigure}{0.32\textwidth}
		\centering
		\includegraphics[width=\textwidth]{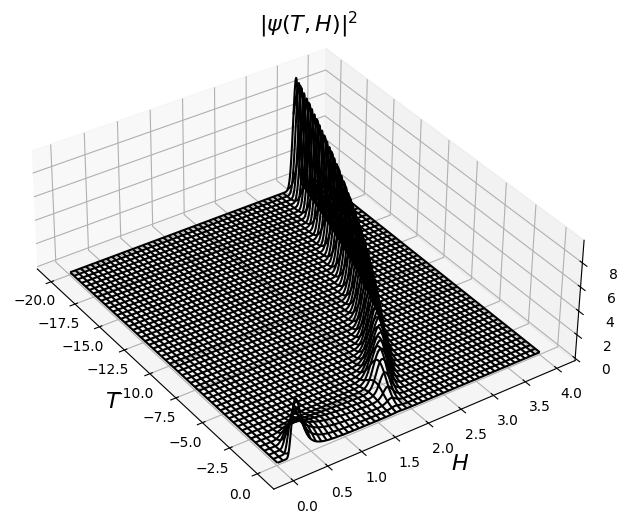}
		\caption{}
		\label{fig:42a}
	\end{subfigure}
	\hfill
	\begin{subfigure}{0.32\textwidth}
		\centering
		\includegraphics[width=\textwidth]{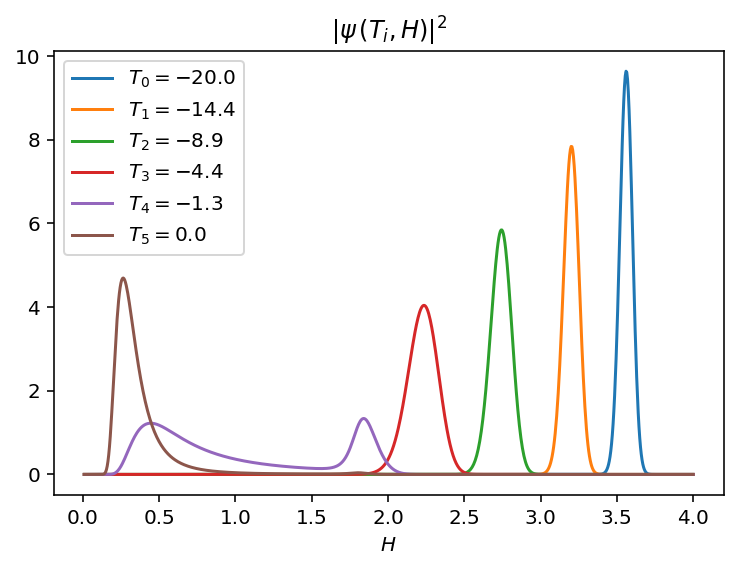}
		\caption{}
		\label{fig:42b}
	\end{subfigure}
	\hfill
	\begin{subfigure}{0.32\textwidth}
		\centering
		\includegraphics[width=\textwidth]{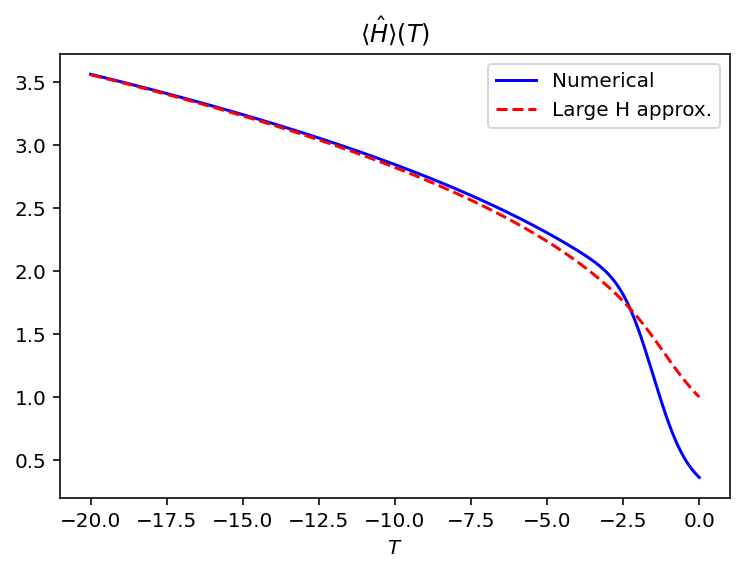}
		\caption{}
		\label{fig:42c}
	\end{subfigure}
	\caption{(a) Square amplitude (\ref{newwave}) for Model 1, showing a well defined trajectory in minisuperspace. (b) Profiles $|\psi(T,H)|^2$ at different values of $T$. (c) Mean value of the Hubble factor (with $T$ as evolution parameter) for the numerical solution compared with approximation (\ref{meanapp}).}
	\label{fig:42}
\end{figure*}

Now, the preceding computations are performed using the large-$H$ approximate wave function (\ref{solp1}). Since we are interested also in the transition from large to small $H$, we use  wave functions such as the one in Fig. \ref{fig:36}, obtained numerically.  We use a finite domain $[0,8]$ for $\xi$ and the Gaussian centered at $\bar{\xi}=4$, with $\sigma=1$. Performing the sum corresponding to (\ref{wavepacket}) and normalizing over $T$, we get the wave packet shown in Fig. \ref{fig:42a}. In Fig. \ref{fig:42b}, we show the profiles of $|\psi(T,H)|^2$ at different times. They show a clear tendency of $\braket{\hat{H}}$ decreasing slowly as $T$ moves forward, until the mean value crosses certain threshold at which the decrease occurs more rapidly.  At remote times, $T<-2.5$ in Fig. \ref{fig:42a}, and at recent times $T>-1$, we have a maximum square amplitude at large and small values of $H$, respectively. In Fig. \ref{fig:42b} we can see the transition from (maximum amplitude at) large to small $H$ occurs at $T\approx -1.3$, where  two local maxima of the square amplitude coexist. 

Further, in Fig. \ref{fig:42c} we compute the mean value $\braket{\hat{H}}(T)$ for the numerical solution and compare with our large-$H$ approximation \ref{meanapp}. If we associate the same geometric meaning to $\braket{H}(T)$ as the Hubble factor, then $\log [a(T)/a_0]=\int_{T_0}^T \braket{H} dT' \equiv N_{e}(T)$, the number of $e$-folds. For the example of Fig. \ref{fig:42c}, we get $N_{e}=53.4$.

\begin{figure*}[ht]
	\centering
	\begin{subfigure}{0.32\textwidth}
		\centering
		\includegraphics[width=\textwidth]{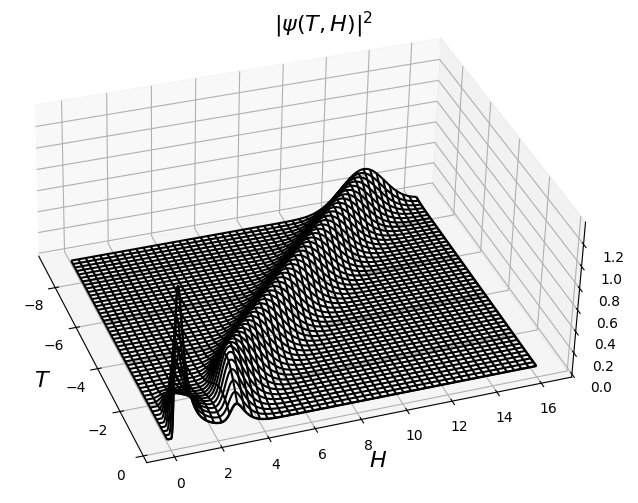}
		\caption{}
		\label{fig:am2}
	\end{subfigure}
	\hfill
	\begin{subfigure}{0.32\textwidth}
		\centering
		\includegraphics[width=\textwidth]{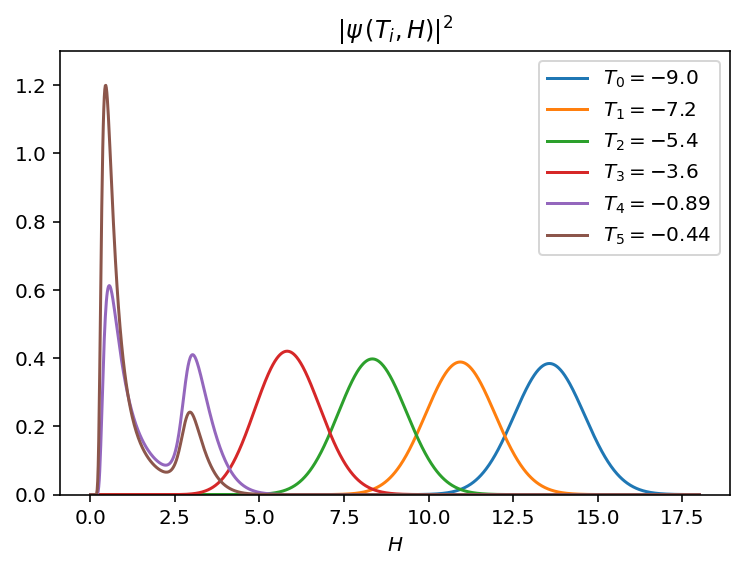}
		\caption{}
		\label{fig:bm2}
	\end{subfigure}
	\hfill
	\begin{subfigure}{0.32\textwidth}
		\centering
		\includegraphics[width=\textwidth]{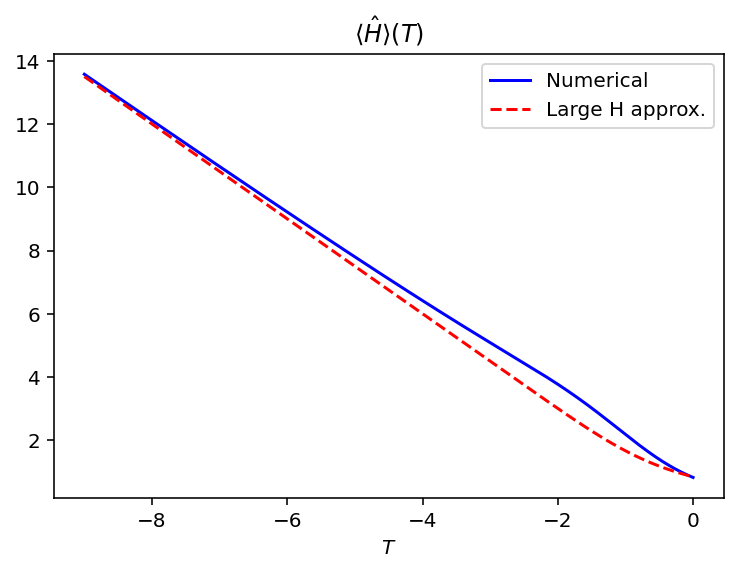}
		\caption{}
		\label{fig:cm2}
	\end{subfigure}
	\caption{(a) Square amplitude (\ref{newwave}) for Model 2. (b) Profiles of $|\psi|^2$ at different values of $T$. (c) Mean value of $H$ obtained numerically.}
	\label{fig:46}
\end{figure*}

Now, for Model 2, using the approximate wave function (\ref{wkbmod2}), we get (integrating again over the full $\xi$-axis) 
\begin{align}
	\Psi(T,H)=\sqrt{\tfrac{2}{3}} e^{-\frac{\sigma^2}{2} (T+\frac{2}{3} L^2 H)^2+i \bar{\xi} (T+\frac{2}{3} L^2 H)},
\end{align} 
whose square norm is    
\begin{align}
	\mathcal{N}^2(T)
	=\frac{1}{\sigma L^2} \int_{\sigma T}^\infty e^{-Y^2} dY=\frac{\sqrt{\pi}}{2 \sigma L^2}  \text{erfc} (\sigma T)
\end{align}
where $Y=\sigma (\frac{2}{3} L^2 H+T)$. Next, mean value of $\hat{H}$ (\ref{meandef}) gives us 
\begin{align}
	\braket{\hat{H}}(T)
	&=\frac{3}{2 L^4 \sigma^2} \int_{\sigma T}^\infty (Y-\sigma T) e^{-Y^2} dY \nonumber \\
	&=\frac{3}{4 L^4 \sigma^2} \big[e^{-\sigma^2 T^2}-\sqrt{\pi } \sigma T \, \text{erfc}(\sigma T)\big]
\end{align}
The normalized wave function (\ref{newwave}), as well as the mean value of $\hat{H}$, obtained numerically are shown in Fig. \ref{fig:46}. For remote times $T<-3$, the shape of $|\psi|^2$ is approximately preserved as $T$ increases, which reflects in the quasi-linear evolution of $\braket{\hat{H}}$.

\subsection{Radiation}\label{sec5.2}
For radiation $w=\frac{1}{3}$, the Hamiltonian constraint (\ref{fullhamil2}) is  
\begin{align}\label{horad}
	H_0=-3 A \frac{F(P)}{f(P)}+\left(\frac{f(P)}{\kappa^2 A}\right)^{\frac{1}{3}}  P_T \approx 0.
\end{align}

For simplicity, instead of the original Hamiltonian constraint (\ref{horad}), we consider the following variant 
 \begin{align}\label{newh0rad}
 	\mathcal{H}\equiv \frac{H_0 A^{1/3}}{[f(P)]^{1/3}}=-G(P) A^{4/3}+\frac{P_T}{\kappa^{2/3}}  \approx 0
 \end{align}
where $G(P)$ is not the same as (\ref{sch1}) but 
\begin{align}\label{grad}
G(P)\equiv	\frac{3 F(P)}{[f(P)]^{4/3}}
\end{align}

Writing the quantum constraint equation $\hat{H} \tilde{\Psi}(T,H)$, with $\hat{H}$ ordered as in (\ref{newh0rad}), using $\hat{A}=i \hbar \p_P$, $P_T=-i \hbar \p_T$, and substituting a separable wave function, $\tilde{\Psi}(T,P)=\exp(i \chi T/\hbar) \Phi(P)$, yields the following $T$-independent equation, 
\begin{align}\label{fractional}
	\left(i \hbar D \right)^{\frac{4}{3}} \Phi(P)=\frac{\chi}{\kappa^{2/3}} \frac{\Phi(P)}{G(P)}	
\end{align}
where $D\equiv \frac{d}{dP}$ and $\chi$ is the separation constant.

\begin{figure}[ht]
	\centering
	\includegraphics[width=.36\textwidth]{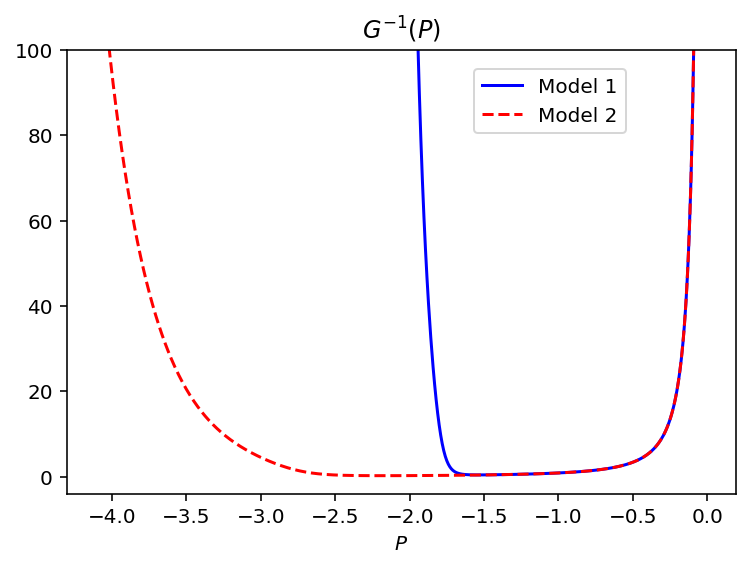}
	\caption{$G^{-1}(P)$ for models (\ref{model1})-(\ref{model2}) with $\la=10^{-6}$ and $L=1$. }
	\label{fig:90}
\end{figure}

Let's first comment on the right-hand side of (\ref{fractional}). Due to the exponent $\frac{4}{3}$, an exponential of positive sign survives in the denominator when considering the asymptotic expansion of (\ref{grad}) as $P\to \infty$, which makes $G^{-1}(P)$ diverge more rapidly than with dust (cf. (\ref{w1m1})-(\ref{gm1app})), specially with Model 1, as can be seen in Fig. \ref{fig:90}. For small $P$, $G^{-1}(P)\approx 2^{4/3}/(3 P^2)$, for both models.

Now, on the left-hand side of (\ref{fractional}) we get $D^{4/3}$. Some possible definitions of this operator are provided by either the Riemann-Liouville (\ref{rl}) or Caputo (\ref{cap}) differential operators of order $\nu \in \mathbb{R}^+$. In Appendix \ref{app3}, we collect some basic results about the intuition behind fractional differential operators and the solution of fractional differential equations (FDEs) based on \cite{diethelm,math6020016}.    

We use Caputos derivative since it allows us to define an initial value problem in the spirit of integer-order differential equations. Since Caputo (and also Riemann-Liouville) derivative (\ref{cap}) is linear, expressing $\Phi(P)$ in terms of its real and imaginary components, (\ref{fractional}) amounts to a system of FDEs of commensurate order (only derivatives of order $\frac{4}{3}$)
\begin{subequations}\label{numrad}
	\begin{align}
		D^{\frac{4}{3}} \text{Re} \Phi&=\frac{\chi}{2 (\hbar^2 \kappa)^{2/3}} \frac{\sqrt{3} \text{Im} \Phi-\text{Re} \Phi}{G(P)}, \\
		D^{\frac{4}{3}} \text{Im} \Phi&=\frac{\chi}{2 (\hbar^2 \kappa)^{2/3}} \frac{-\text{Im} \Phi-\sqrt{3} \text{Re} \Phi}{G(P)},
	\end{align}
\end{subequations}
which must be supplemented with initial conditions $\Phi(P_0)$ and $\Phi'(P_0)$.

To solve (\ref{numrad}) we resort directly to the numerical methods available in the literature, in particular the Matlab routine \verb|FDE_PI1_Ex| \cite{math6020016} and its  implementation in Julia language \verb|PIEX| \cite{qu2025fractionaldiffeqjl}. 

\begin{figure*}[ht]
	\centering
	\begin{subfigure}{0.32\textwidth}
		\centering
		\includegraphics[width=\textwidth]{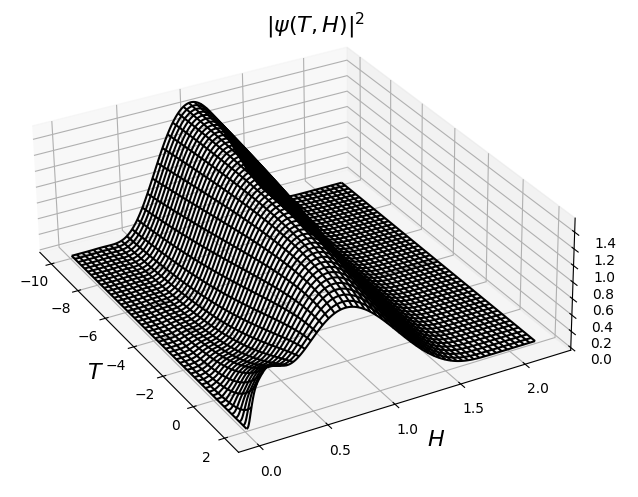} 
		\caption{}
		\label{}
	\end{subfigure}
	\hfill
	\begin{subfigure}{0.32\textwidth}
		\centering
		\includegraphics[width=\textwidth]{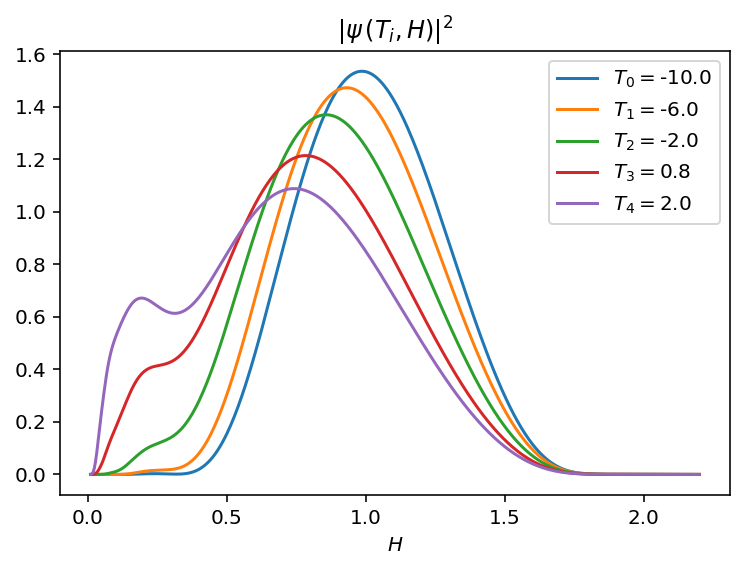}
		\caption{}
		\label{}
	\end{subfigure}
	\hfill
	\begin{subfigure}{0.32\textwidth}
		\centering
		\includegraphics[width=\textwidth]{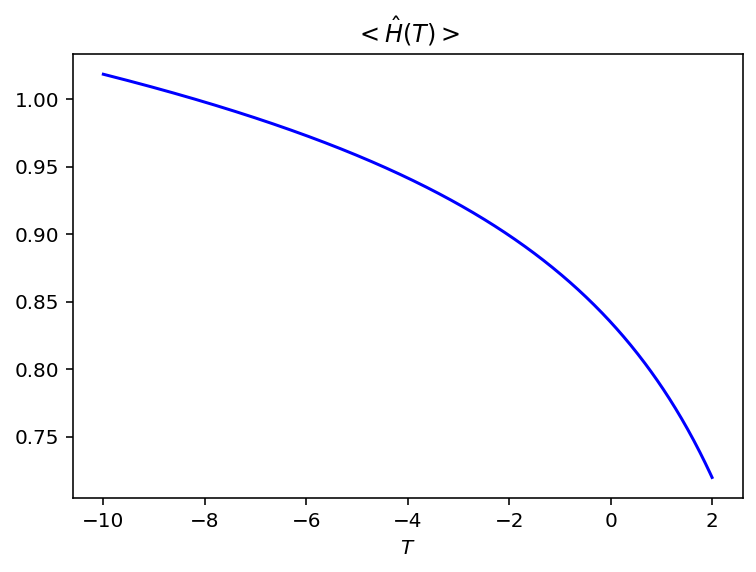} 
		\caption{}
		\label{}
	\end{subfigure}
	\caption{Numerical solution for Model 1 with $\bar{\chi}=4$ and $\sigma=0.5$. We obtain $N_e=11.0$}
	\label{fig:48}
\end{figure*}

\begin{figure*}[ht]
	\centering
	\begin{subfigure}{0.32\textwidth}
		\centering
		\includegraphics[width=\textwidth]{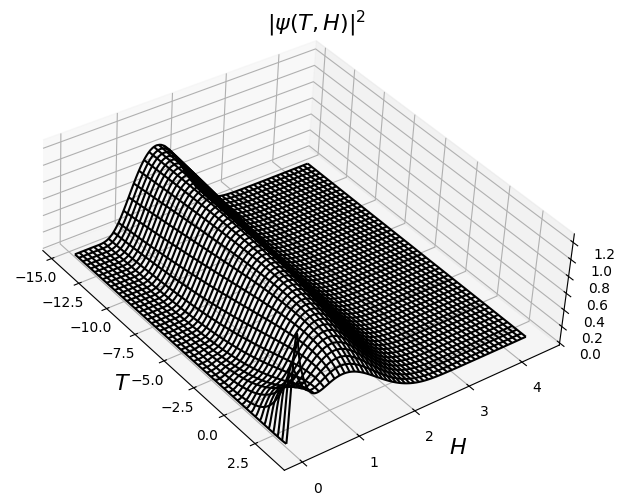} 
		\caption{}
		\label{}
	\end{subfigure}
	\hfill
	\begin{subfigure}{0.32\textwidth}
		\centering
		\includegraphics[width=\textwidth]{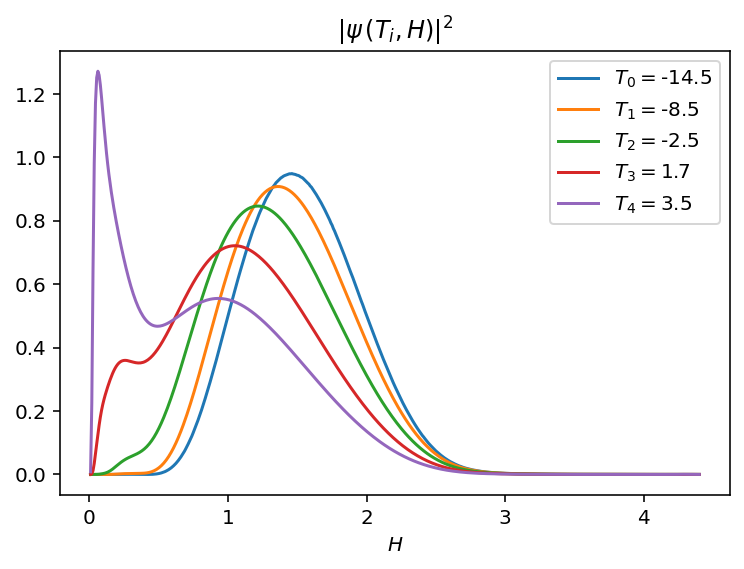}
		\caption{}
		\label{}
	\end{subfigure}
	\hfill
	\begin{subfigure}{0.32\textwidth}
		\centering
		\includegraphics[width=\textwidth]{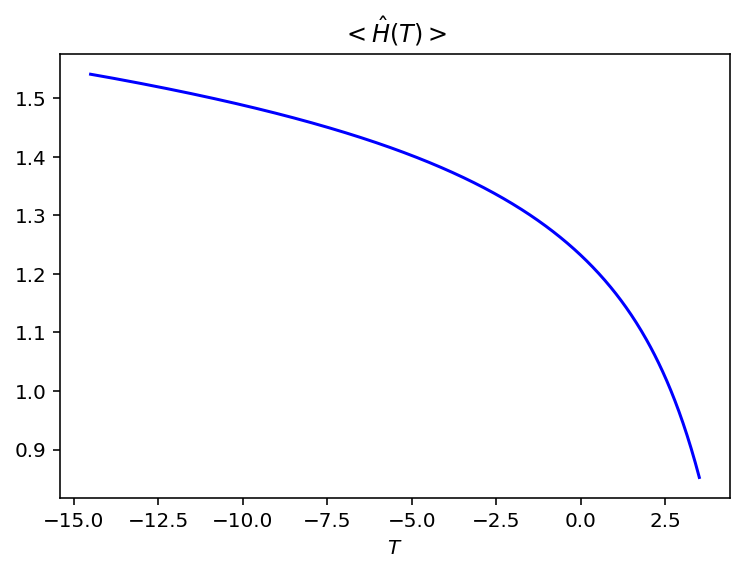} 
		\caption{}
		\label{}
	\end{subfigure}
	\caption{Numerical solution for Model 2 with $\bar{\chi}=4$ and $\sigma=0.5$. We obtain $N_e=24.5$}
	\label{fig:49}
\end{figure*}

We perform the process as with dust. Compute $\Phi(0,P)$ solving (\ref{numrad}) numerically for discrete values $\chi$ in the range $[0,8]$. Next, evolve with the phase $e^{i \chi T/\hbar}$ and perform the sum corresponding to (\ref{wavepacket}) with a Gaussian distribution centered at $\bar{\chi}=4$ with $\sigma=0.5$. Our numerical results are presented in Fig. \ref{fig:48} for Model 1 and Fig. \ref{fig:49} for Model 2. 

\begin{figure}[t]
	\centering
	\includegraphics[width=0.34\textwidth]{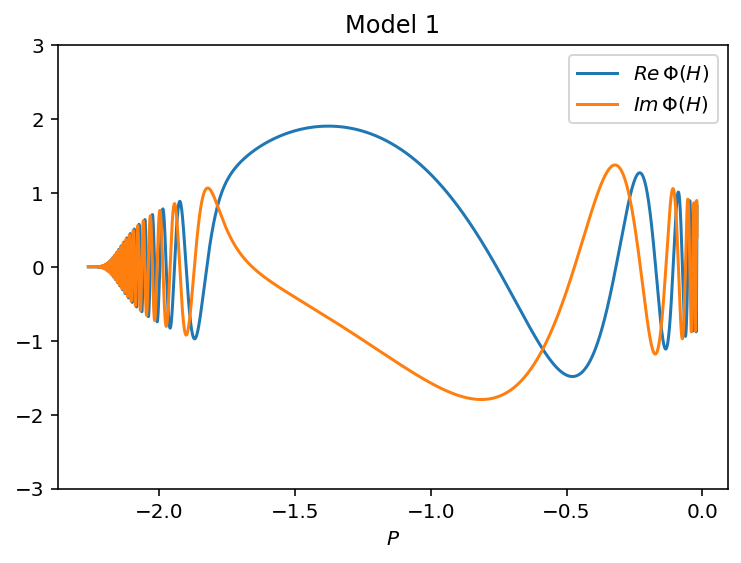}
	\hfill
	\includegraphics[width=0.36\textwidth]{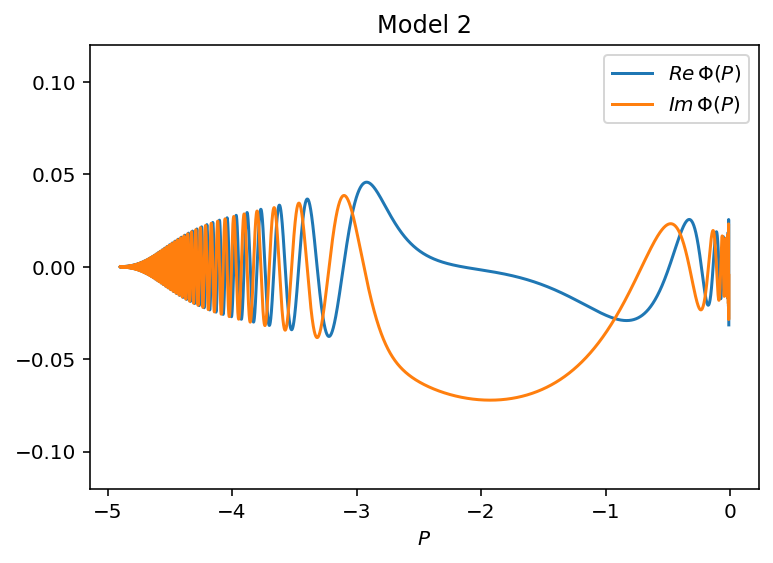}
	\caption{Numerical solution of (\ref{numrad}) with $\chi=4$, step size $dP=10^{-4}$ and initial conditions $\Phi(P_0)=0$, $\Phi'(P_0)=10^{-2}$. $P_0$ is chosen according to Fig. \ref{fig:90}.}
	\label{fig:38}
\end{figure}

\section{Conclusions}\label{sec6}
We studied the Wheeler-DeWitt quantization minisuperspace FRW models corresponding to the Geometric Inflation (GI) proposal, which belongs itself to a class of higher-curvature theories called quasi topological gravity. 

First, we construct an effective 1D FRW Lagrangian (\ref{frwgeometric}) from which we can obtain the modified Friedmann equations of GI. Although such Lagrangian depends on the second time derivative of the scale factor, it does so linearly and the resulting equation of motion is not fourth but second order. In other words, it is a constrained higher-derivative system. Next, we perform the Hamiltonian formulation of the FRW GI Lagrangian, using  Ostrogradsky's construction and Dirac's algorithm constrained systems. Since our  Dirac's bracket $\{a,H\}_D$ is given by a nontrivial function of phase space, we look for an actual canonical pair. Although it would have been preferable to work with the scale factor $a$, its canonical momentum $p_a$ is generally a non-polynomial function of $(a,H)$, which complicates obtaining $H_0(a,p_a)$ explicitly. An alternative is working with $P=-H$ and its conjugate momentum $A(a,H)$ (\ref{AP}). All of this amounts to a generalization of the Hamiltonian formulation of the  cubic model \cite{einsteiniancubic} to the full tower of curvature invariants.

Second, we implement Dirac quantization scheme by which we postulate the wave function $\Psi(T,A)$, to be annihilated by the Hamiltonian constraint operator. The matter part is a perfect fluid with barotropic linear equation of state, for which a Hamiltonian description is provided by Schutz formalism.  In position space the WDW equation is of infinite order in derivatives of $A$, and thus we only consider a first-order WKB approximation. Since $A(a,H)$ has no direct physical interpretation (except for the ordinary GR case, in which $A\propto a^3$), the momentum representation it worth considering, in which case  $\tilde{\Psi}=\tilde{\Psi}(T,H)$. Since the perfect fluid's momentum appears linearly in the Hamiltonian constraint, the WDW equation can be taken to a form resembling the time-dependent Schrodinger equation that admits separation of variables.  For dust $w=0$, the $T$-independent equation is a simple first order equation, with solutions labeled by a non-negative continuous parameter $\xi$. By using a  Gaussian distribution $\mu(\xi)$, we construct wave packet $\tilde{\Psi}(T,H,\xi)$ tracing a marked trajectory in minisuperspace $(T,H)$. In view of the $T$-dependent WDW equation, we choose to treat $T$ as external parameter by defining an effective wave function $\psi(T,H)$ that is normalized at every $T$. We use this setting to compute the mean value of the quantum Hubble factor $\hat{H}$. For Model 1, our numerical computations reveal a behavior resembling a classical inflationary solution, with a transient stage of approximately constant $\braket{H}(T)$ until it starts decaying more rapidly, marking the end of inflation.

On the other hand, with radiation $w=1/3$, we face a significant complication. The $T$-independent WDW equation in the momentum representation involves a differential operator of fractional order. We use Caputo derivative since it allows to define an initial value problem similar to  ordinary differential equations. The behavior of the numerical solutions is slightly more delicate than in the case of dust, but we still manage to construct wave packets that yield a mean value of $\hat{H}$ not too far from a classical inflationary solution, in particular with Model 2, since the potential in this case diverges more slowly than with Model 1.        

There are several aspects about the quantum cosmology of geometric inflation such as the boundary condition that solutions of the GI WDW equations must satisfy. The path integral formulation of the wave function and its connection with our WDW solutions. A thorough analysis of the fractional differential equations for radiation and looking for some analytic solutions or at least some approximations that back up our numerical results. The quantization of the spatially closed FRW model was not addressed here since we need to find another suitable canonical pair or try a different approach such as path integral quantization \cite{Mondal2025}. 

\appendix
\section{Perfect fluid}\label{app1}
Here we briefly summarize Schutz velocity potential representation of a perfect fluid following references \cite{PhysRevD.2.2762,PhysRevD.4.3559} and its FRW specialization \cite{Lapchinskii1977,lemos,Burton-Villalobos2025}. Let's consider a single component perfect fluid in the frame of baryon number conservation. The total rest mass of a sample containing $N_B$ baryons is $M_0=m_0 N_B$, all other forms of energy are contained in the internal energy. Thus, the total fluid's mass-energy is $E=U+M_0$ or, in terms of densities, 
\begin{align}\label{onepluspi}
	\rho=\frac{M_0+U}{V} =\rho_0+u\equiv \rho_0 (1+\Pi).
\end{align}
where $\rho_0$ is the density of rest mass and $\Pi \equiv \frac{u}{\rho_0}=\frac{U}{M_0}$ is the specific internal energy. 

According to the First Law of Thermodynamics, the energy per unit rest mass, $\delta q$, added to the fluid in a quasistatic process preserving baryon number is 
\begin{align}
	\delta q\equiv \frac{\delta Q}{M_0}	= \frac{\delta U}{M_0}+pd \frac{V}{M_0}=\delta \Pi+p \delta \left(\frac{1}{\rho_0}\right),
\end{align}
where $p$ is the fluid's pressure.
 
Assuming an equation of state of the form $p=p(\rho_0, \Pi)$, there are two functions of $\rho_0$ and $\Pi$, called the specific entropy $S$ and the temperature $T$, such that $\delta q=TdS$, that is,
\begin{align}\label{imagin}
	TdS=d \Pi+pd \left( \frac{1}{\rho_0}\right)
\end{align}

On the other hand, the specific inertial mass or specific enthalpy $\mu$ is defined as follows 
\begin{align}\label{inermas}
	\mu &\equiv	\frac{\rho+p}{\rho_0} =1+\Pi+\frac{p}{\rho_0}, 
\end{align}
then, $d \mu=d \Pi+\frac{dp}{\rho_0}+p d (\frac{1}{\rho_0})$ and we can rewrite (\ref{imagin}) as 
\begin{align}\label{usvn}
	\rho_0 d \mu-\rho_0 T dS=dp,
\end{align}
showing that the equation of state can be expressed in the form 
\begin{align}\label{eqofstate}
	p=p(\mu, S)
\end{align}

For the linear barotropic equation, $p=w \rho$, and using (\ref{onepluspi}), we can rewrite (\ref{imagin}) as follows \cite{Lapchinskii1977}
\begin{align}
	TdS&=d\Pi-\frac{w \rho}{\rho_0^2} d \rho_0=d\Pi-\frac{w (1+\Pi)}{\rho_0} d \rho_0 \nonumber \\
	&=(1+\Pi) d [\ln (1+\Pi)-w \ln \rho_0], 
\end{align}
which yields \cite{Lapchinskii1977} $S=\ln (1+\Pi)-w \ln \rho_0 =\ln \frac{\rho}{\rho_0}-w \ln \rho_0$. Therefore, 
\begin{align}\label{isab}
\rho=\rho_0^{1+w} e^{S}
\end{align}
%
%

\subsection{FRW fluid's action}
In Schutz's formalism \cite{PhysRevD.2.2762,PhysRevD.4.3559}, the fluid's  4-velocity is defined in terms of six potentials. In the FRW case we only require the following \cite{Lapchinskii1977} 
\begin{align}\label{velpot}
	U_m=\frac{1}{\mu} (\p_m \phi+\theta \p_m S)
\end{align}

From the normalization of the 4-velocity, $U^m U_n=-N^{-2} U_0^2=-1$, we get $U_0=N$ and,
\begin{align}\label{mu}
\mu=\frac{\p_t \phi+\theta \p_t S}{N}=\frac{\dot{\phi}+\theta \dot{S}}{N},
\end{align}
where we use the FRW metric (\ref{frwmetric}).

Schutz action for a perfect fluid with equation of state (\ref{eqofstate}) is $S_{\text{fluid}}\equiv \int d^4 x \sqrt{-g} \, p(\mu, S)$. Thus, in the FRW case we have
\begin{align}
	L_{\text{fluid}}=N a^3 p(\mu, S)
\end{align}
First, the momenta conjugate to $\phi, \theta, S$ are, respectively,   
\begin{subequations}
	\begin{align}
		p_\phi&=\frac{\p L_{\text{fluid}}}{\p \dot{\phi}}=a^3 \frac{\p p}{\p \mu} \\
		p_\theta&=\frac{\p L_{\text{fluid}}}{\p \dot{\theta}}=0, \\
		p_S&=\frac{\p L_{\text{fluid}}}{\p \dot{S}}=a^3 \frac{\p p}{\p \mu} \theta.
	\end{align}
\end{subequations}

Using (\ref{mu}), the canonical Hamiltonian is
\begin{align}
	H_{\text{fluid}}& \equiv \dot{\phi} p_\phi+\dot{\theta} p_\theta+\dot{S} p_S-L_{\text{fluid}} \nonumber \\
	&=N a^3 \left(\mu \frac{\p p}{\p \mu}-p \right)=N a^3 \rho, 
\end{align}
where we use $\rho_0=\frac{\p p}{\p \mu}$, from (\ref{usvn}), and then (\ref{inermas}).  

For a linear barotropic perfect fluid, using (\ref{isab}) and $p_\phi=a^3 \rho_0$, we obtain 
\begin{align}\label{hamilfluid}
	H_{\text{fluid}}=N \frac{p_\phi^{1+w}}{a^{3 w}} e^{S}
\end{align}

The total fluid Hamiltonian is $H_{\text{fluid}}^{\text{tot}}=N \frac{p_\phi^{1+w}}{a^{3 w}} e^{S}+c_1 \chi_1+c_2 \chi_2$, where $c_1$ and $c_2$ are multipliers of the two primary constraints, 
\begin{align}\label{constfluid}
\begin{split}
\chi_1&=p_\theta \approx 0, \\
\chi_2&=p_S-\theta p_\phi \approx 0.
\end{split}
\end{align}

Performing Dirac's algorithm, we identify the multipliers $c_2=0$ and $c_1=-N \frac{p_\phi^w}{a^{3 w}} e^{S}$ and the constraints (\ref{constfluid}) are second class. The nonzero basic Dirac brackets are given by 
\begin{align}
	\{\phi,p_\phi\}_D=1, && \{s,p_S\}_D=1, && \{S, \theta\}_D=p_\phi^{-1}
\end{align}
Since $p_\theta$ has vanishing Dirac bracket with all other canonical variables, we set $p_\theta=0$, and $\theta=\frac{p_S}{p_\phi}$. Thus, our independent canonical variables are $\phi,p_\phi,S,p_S$. 

Solving the second class constraints we are left with Hamiltonian (\ref{hamilfluid}).  The following canonical transformation \cite{lemos} ($T$ not related to temperature anymore) 
\begin{align}
	T=-p_S e^{-S} p_\phi^{-(1+w)}, && P_T=p_\phi^{1+w} e^S, \\
	\bar{\phi}=\phi+(1+w) \frac{p_S}{p_\phi}, && \bar{p}_\phi=p_\phi
\end{align}
yields 
\begin{align}\label{hfluid}
	H_{\text{fluid}}=N \frac{P_T}{a^{3w}}
\end{align}
Note that the new configuration space variable $T$ has monotonic evolution
\begin{align}\label{derit}
\dot{T}=\{T,H_{\text{fluid}}\}=\frac{N}{a^{3 w}}
\end{align}

\section{Kummer confluent hypergeometric function}\label{app2}
Define $x_0=\sigma T$, then the mean value (\ref{specialinte}) is proportional to $J_{4/3}(x_0)$, where 
\begin{align}
	J_\nu(x_0)\equiv \int_0^\infty u^{\nu-1} e^{-u^2-2 x_0 u} du
\end{align}
Using $e^{-2 x_0 u}=\cosh (2 x_0 u)-\sinh (2x_0 u)$, we have 
\begin{align}\label{decompose}
	J_\nu(x_0)&=\int_0^\infty u^{\nu-1} e^{-u^2} \cosh (2 x_0 u) du \nonumber \\ 
	& - \int_0^\infty u^{\nu-1} e^{-u^2} \sinh (2x_0 u) du \nonumber \\ 
	& \equiv J_\nu^{(+)}(x_0)-J_\nu^{(-)}(x_0)
\end{align}
Now, substituting the series expansion of hyperbolic functions, $\cosh (2 x_0 u)=\sum_{k=0}^\infty (2 x_0 u)^{2k}/(2k)!$, $\sinh (2 x_0 u)=\sum_{k=0}^\infty  (2 x_0 u)^{2k+1}/(2k+1)!$, and integrate order by order

\begin{align}
	J_\nu^{(+)}(x_0)&=\sum_{k=0}^\infty \frac{(2 x_0)^{2k} }{(2k)!} \int_0^\infty u^{2k+\nu-1} e^{-u^2} du \nonumber \\ 
	&=\sum_{k=0}^\infty \frac{(2 x_0)^{2k} }{(2k)!} \frac{1}{2} \Gamma \left(\frac{\nu}{2}+k\right) \nonumber \\ 
	&=\frac{1}{2} \Gamma \left(\frac{\nu}{2} \right) \sum_{k=0}^\infty \frac{(2 x_0)^{2k} }{(2k)!}  \left(\frac{\nu}{2} \right)_k \nonumber \\
	&=\frac{1}{2} \Gamma \left(\frac{\nu}{2} \right) \sum_{k=0}^\infty \frac{(x_0)^{2k} }{k! (\frac{1}{2})_k}  \left(\frac{\nu}{2} \right)_k,
\end{align}
using $(2k)!=4^k k! (\frac{1}{2})_k$ ($(\, )_k$ denotes the Pochhammer symbol). The last series corresponds to the Kummer confluent hypergeometric function \cite{kummer} 
\begin{align}
	J_\nu^{(+)}(x_0)=\frac{1}{2} \Gamma \left(\frac{\nu}{2} \right)  \, _1 F_1\left( \frac{\nu}{2}; \frac{1}{2},x_0^2\right) 
\end{align}

Similarly, 
\begin{align}
	J_\nu^{(-)}(x_0)&=\sum_{k=0}^\infty \frac{(2 x_0)^{2k+1}}{(2k+1)!} \int_0^\infty u^{2k+1+\nu-1} e^{-u^2} du \nonumber \\
	&=\sum_{k=0}^\infty \frac{(2 x_0)^{2k+1}}{(2k+1)!} \frac{1}{2} \Gamma \left(\frac{1+\nu}{2}+k\right)  \nonumber \\
	&=\frac{1}{2} \Gamma \left(\frac{\nu+1}{2}\right) \sum_{k=0}^\infty \frac{2^{2k+1} (x_0)^{2k+1}}{(2k+1)!}   \left(\frac{\nu+1}{2} \right)_k  \nonumber \\
	&=x_0 \Gamma \left(\frac{\nu+1}{2}\right) \sum_{k=0}^\infty \frac{(x_0)^{2k}}{k! (\frac{3}{2})_k}   \left(\frac{\nu+1}{2} \right)_k \nonumber \\
	&=x_0 \Gamma \left(\frac{\nu+1}{2}\right)   \, _1 F_1\left( \frac{\nu+1}{2}, \frac{3}{2}, x_0^2\right),
\end{align}
using $(2k+1)!=4^k k! (\frac{3}{2})_k$. Therefore
\begin{align}
	J_\nu(x_0)=\frac{1}{2} \Gamma \left(\tfrac{\nu}{2} \right)  \, _1 F_1\left(\tfrac{\nu}{2}; \tfrac{1}{2},x_0^2\right) - x_0 \Gamma \left(\tfrac{\nu+1}{2}\right)   \, _1 F_1\left( \tfrac{\nu+1}{2}, \tfrac{3}{2}, x_0^2\right) 
\end{align}
and for (\ref{specialinte}), $\nu=\frac{4}{3}$, $J_{4/3}(x_0)= \Gamma (\frac{2}{3})  \, _1 F_1(\frac{2}{3}; \tfrac{1}{2},x_0^2) - x_0 \Gamma (\frac{7}{6}) \, _1 F_1(\frac{7}{6}, \frac{3}{2}, x_0^2)$.

\section{Fractional derivatives}\label{app3}
The following is a collection of basic results about fractional derivatives and fractional differential equations based on \cite{diethelm}. 

Let $y(x)$ be a continuous function in the compact subset $[x_0,x_1]$ of $\mathbb{R}$. According to the Fundamental Theorem of Calculus, if $Y(x)\equiv \int_{x_0}^x y(t) dt$, then $Y$ is differentiable and $Y'(x)\equiv \frac{dY(x)}{dx}=y(x)$. Let's denote the ordinary derivative operator by $D$, that is, $D y(x)\equiv \frac{dy(x)}{dx}=y'(x)$. On the other hand, the primitive of $y(x)$, centered at $x_0$, is defined as $J_{x_0} y(x)\equiv \int_{x_0}^x y(t) dt$ ($x_0 \le x\le x_1$). Further, for $n\in \mathbb{N}$ and $n\ge 2$, the $n$-fold iterations $D^n$ and $J_{x_0}^n$ are defined through recursion, 
\begin{align}
	D^n\equiv D D^{n-1}, && J_{x_0}^n=J_{x_0} J_{x_0}^{n-1}.
\end{align}
By the fundamental theorem, which can be written as $D J_{x_0} y=y$, it is readily shown that $D^n J_{x_0}^n y=y$.

Using Fubini’s theorem and mathematical induction, it can be shown that, if $y(x)$ is Riemann integrable in $[x_0,x_1]$, then 
\begin{align}\label{nfoldintegral}
	J_{x_0}^n y(x)=\frac{1}{(n-1)!} \int_{x_0}^x (x-t)^{n-1} y(t) dt.
\end{align}
Further, if $y$ has a continuous $n$-th derivative on $[x_0,x_1]$, and $m,n\in \mathbb{N}$, with $m>n$, then
\begin{align}\label{derivinte}
	D^n y=D^m J_{x_0}^{m-n} y.
\end{align}
Expressions (\ref{nfoldintegral}) and (\ref{derivinte}) lend themselves to defining integration and differentiation operators of non-integer degree.

If $\nu \in \mathbb{R}^+$ and $y \in L_1[x_0,x_1]$ (the space of Lebesgue integrable functions of order 1), the Riemann-Liouville fractional integral operator of order $\nu$ is defined by \cite{diethelm}  
\begin{align}
	J_{x_0}^\nu y(x)\equiv \frac{1}{\Gamma(\nu)} \int_{x_0}^x (x-t)^{\nu-1} y(t) dt.
\end{align}
For $\nu=0$, $J_{x_0}^0\equiv I$, the identity operator.

Now, let $\nu \in \mathbb{R}^+$ and $\mu=\lceil \nu \rceil$ (the ceiling of $\nu$), the Riemann-Liouville (R-L) derivative of order $\nu$, $D_{x_0}^\nu$, is defined by  
\begin{align}\label{rl}
	D_{x_0}^\nu y(x)&\equiv D^\mu J_{x_0}^{\mu-\nu} y(x) \nonumber \\ 
	&=D^\mu \left( \frac{1}{\Gamma(\mu-\nu)} \int_{x_0}^x (x-t)^{\mu-\nu-1} y(t) dt\right) 
\end{align}
For $\nu=0$, $D_{x_0}^0\equiv I$, the identity operator.

An alternative definition is known as Caputo derivative, $\hat{D}_{x_0}^\nu$. Let $\nu \ge 0$,  $\mu=\lceil \nu \rceil$ and $D^\mu y\in L^1[a,b]$, then 
\begin{align}\label{cap}
	\hat{D}_{x_0}^\nu y(x)&\equiv J_{x_0}^{\mu-\nu} [D^\mu y(x)] \nonumber \\ 
	&=\frac{1}{\Gamma(\mu-\nu)} \int_{x_0}^x (x-t)^{\mu-\nu-1} D^\mu y(t) dt 
\end{align}

For $\nu \ge 0$, Caputo (\ref{cap}) and Riemann-Liouville (\ref{rl}) derivatives are related as follows \cite{diethelm}, 
\begin{align}
	\hat{D}_{x_0}^\nu y=D_{x_0}^\nu \left(y-T_{\mu-1}[y; x_0]\right) 
\end{align}
where $T_{\mu-1} [y; x_0]$ is the Taylor polynomial of degree $\mu-1$ centered at $x_0$, 
\begin{align}
	T_{\mu-1}[y; x_0] (x)=\sum_{k=0}^{\mu-1} \frac{y^{(k)}(x_0)}{k!} (x-x_0)^k.
\end{align}
For $\mu=0$, $T_{\mu-1} [y; x_0] \equiv 0$.

A remarkable difference between R-L and Caputo derivatives is their behavior with respect to constant functions. Using definition (\ref{cap}) we obtain, for $\nu \notin \mathbb{N}$, $\hat{D}_{x_0}^\nu 1=0$, whereas using (\ref{rl}), we get $D_{x_0}^\nu 1=\frac{1}{\Gamma(1-\nu)} (x-x_0)^{-\nu}$. Therefore, the Caputo derivative allows to define an initial value problem for a fractional differential equation (FDE)
\begin{align}\label{fraceq}
	\hat{D}_{x_0}^\nu y(x)=F(x,y(x)), 
\end{align} 
where $F$ is a continuous function, with the usual initial conditions $y(x_0)=y_0$,  $y'(x_0)=y_0^{(1)}$, up to $y^{(\mu-1)}(x_0)=y_0^{(\mu-1)}$.

Although both $D_{x_0}^\nu$ and $\hat{D}_{x_0}$ provide left inverses of the R-L fractional integral $J_{x_0}^\nu$, neither of them is a right inverse. In particular,  
\begin{align}\label{relat}
	J_{x_0}^\nu \hat{D}^\nu_{x_0} y(x)
	=y(x)-T_{\mu-1}[y; x_0] (x)
\end{align}

Applying $J_{x_0}^\nu$ to both sides of (\ref{fraceq}) and using (\ref{relat}) yields the following  Volterra Integral Equation (VIE) \cite{diethelm,math6020016} 
\begin{align}\label{vie}
	y(x)=T_{\mu-1}[y;x_0](x)+\frac{1}{\Gamma(\nu)} \int_{x_0}^x (x-t)^{\nu-1} F(t, y(t)) dt
\end{align}
(\ref{vie}) exposes the non-local character of fractional differential equations; $y(x)$ depends on the previous values $y(x')$ with $x'<x$.  

Now, in (\ref{numrad}) we have a system of fractional differential equations (FDEs) of the form 
\begin{align}\label{systemfde}
\begin{split}
D_{x_0}^{\nu_1} y_1(x)=F_1(x,y_1,y_2), \\
D_{x_0}^{\nu_2} y_2(x)=F_2(x,y_1,y_2),
\end{split}
\end{align}
with $\nu_1=\frac{4}{3}=\nu_2$ (this type of system of FDEs with a single fractional order is called commensurate). Since $\lceil \frac{4}{3} \rceil=2$, we must supplement with initial conditions $y_i|_{x_0}$ and $\frac{dy_i}{dx}|_{x_0}$.

System (\ref{systemfde}) of FDEs is equivalent to a system of VIEs, 
\begin{align}
\begin{split}
y_1(x)=T_{\mu_1-1}[y_1;x_0](x)+\frac{1}{\Gamma(\nu_1)} \int_{x_0}^x (x-t)^{\nu_1-1} F_1(t, y) dt, \\
y_2(x)=T_{\mu_2-1}[y_2;x_0](x)+\frac{1}{\Gamma(\nu_2)} \int_{x_0}^x (x-t)^{\nu_2-1} F_2(t, y) dt	
\end{split}
\end{align}

As explained in \cite{math6020016}, numerical methods to solve FDEs made use of the VIE formulation. First of all, the integral in (\ref{vie}) is replaced by a finite sum over a grid $x_n=x_0+n h$, with constant step-size $h>0$ and $n\in \mathbb{N}$, 
\begin{align}
	y(x_n)=T_{\mu-1}[y;t_0]+\frac{1}{\Gamma(\nu)} \sum_{j=0}^{n-1} \int_{t_j}^{t_{j+1}} (t_n-\tau)^{\nu-1} F(\tau,y(\tau)) d\tau
\end{align}

For the Explicit Product Integration (PI) rectangular method, the integral is approximated as \cite{math6020016}
\begin{align}
	y(x_n)=T_{\mu-1}[y;t_0](t_n)+h^\nu \sum_{j=0}^{n-1} b_{n-j-1}^{\nu} F(t_j,y_y)
\end{align}
with $b_n^{\nu}=\frac{(n+1)^\nu-n^\nu}{\Gamma(\nu+1)}$.

\centerline{\bf Acknowledgements}
N. E. Martínez Pérez thanks SECIHTI for financial support.

\bibliography{biblio}

\end{document}